\documentclass{aa}

\usepackage{amssymb}
\usepackage{xcolor,subcaption}
\usepackage[export]{adjustbox}
\usepackage{flushend}
\usepackage[utf8]{inputenc}

\usepackage{supertabular}

\usepackage{hyperref}

\usepackage[compact]{titlesec}  
\titlespacing{\section}{0pt}{12pt}{7pt}
\titlespacing{\subsection}{0pt}{9pt}{4pt}

\newcommand{\frb}{FRB~20180916B}

\newcommand{\urm}{\,rad\,m$^{-2}$}

\newcommand\rss{FRB~20201124A}
\newcommand\RGC{FRB~20200120E}
\newcommand{\rAO}{FRB~20121102A}
\newcommand{\rfast}{FRB~20180301A}

\usepackage[T1]{fontenc}

\DeclareRobustCommand{\VAN}[3]{#2}
\let\VANthebibliography\thebibliography
\def\thebibliography{\DeclareRobustCommand{\VAN}[3]{##3}\VANthebibliography}

\usepackage{graphicx}	% Including figure files
\usepackage{subcaption} % for them sub-floats
\usepackage{booktabs}   % beautiful tables
\usepackage{amsmath}	% Advanced maths commands
\usepackage{amssymb}	% Extra maths symbols
\usepackage{newtxtext,newtxmath}
\usepackage{multirow}
\usepackage{multicol}
\newcommand{\cmd}[1]{\textsc{#1}}
\newcommand{\painf}{PA$_\infty$}
\newcommand{\epainf}{$\langle \mathrm{PA}_\infty \rangle$}

\newcommand{\gcone}{$\mathbf{G}_1$}
\newcommand{\gctwo}{$\mathbf{G}_2$}
\newcommand{\oone}{$\mathbf{O}_1$}
\newcommand{\otwo}{$\mathbf{O}_2$}
\newcommand{\othree}{$\mathbf{O}_3$}

\defcitealias{21DongziPA}{DL21}
\defcitealias{21PastorR3}{PM21}
\defcitealias{23McKinvenR3}{RM23}
\defcitealias{24GopinathR3}{AG23}
\defcitealias{24BethapudiRM}{SB24}
\defcitealias{Landau1976Mechanics}{LL}
\graphicspath{{figures/}}
\title{Constraining the origin of the long term periodicity of FRB~20180916B with Polarization Position Angle}
\titlerunning{\frb~PA}

\author{
S. Bethapudi\inst{\ref{i:bonn}}\thanks{sbethapudi@mpifr-bonn.mpg.de} \and D. Z. Li\inst{\ref{i:princeton}} \and L. G. Spitler\inst{\ref{i:bonn}} \and V. R. Marthi\inst{\ref{i:pune}} \and M. L. Bause\inst{\ref{i:bonn}} \and R. A. Main\inst{\ref{i:mcgill},\ref{i:tsi},\ref{i:bonn}} \and R. S. Wharton\inst{\ref{i:bonn}}.
}
\institute{
Max-Planck-Institute for Radio Astronomy, Auf dem H{\"u}gel 69, Bonn, Germany, 53121\label{i:bonn} \and
Department of Astrophysical Sciences, Princeton University, Princeton, NJ 08544, USA\label{i:princeton} \and
National Centre for Radio Astrophysics, Ganeshkhind, Tata Institute of Fundamental Research, Post Bag~3, Pune - 411~007, India.\label{i:pune} \and
Department of Physics, McGill University, 3600 rue University, Montr\'{e}al, QC H3A 2T8, Canada\label{i:mcgill} \and
Trottier Space Institute, McGill University, 3550 rue University, Montr\'{e}al, QC H3A 2A7, Canada\label{i:tsi} }

\abstract{FRB~20180916B is a repeating Fast Radio Burst (FRB) which produces bursts in a 5.1 day active window which repeats with a 16.34 day period.
Models have been proposed to explain the periodicity using dynamical phenomena such as rotation, precession or orbital motion.
Polarization Position Angle (PA) of the bursts can be used to distinguish and constraint the origin of the long term periodicity of the FRB.
}%
{
We aim to study the PA variability on short (within an observation) and long timescales (from observation to observation). 
Given the periodicity of the source, we also study the PA variations within the active window and across multiple windows.
Then, we aim to compare the observed PA variability with the predictions of various dynamical progenitor models for the FRB.
}%
{
We use the calibrated burst dataset detected by uGMRT in Band~4 (650 MHz) which have been published in \citet{24BethapudiRM}.
We transform the PA measured at 650 MHz to infinite frequency such that PAs measured in different observations are consistent, and finally measure the changes within and across active windows.
}%
{
We find that PA of the bursts vary according to the periodicity of the source.
We constrain the PA variability to be within seven degrees on timescales less than four hours for all MJDs.
We also tentatively find that the PA varies within an active window with a variability of few degrees per hour. 
In addition, we also tentatively note the PA measured at the same phase in the active window varies from one cycle to another.
}%
{
Using the findings, we constrain rotational, precession and binary progenitor models.
Rotational model partially agrees with observed PA variability but requires further study to fully constrain.
We robustly rule out all flavors of precessional models where either precession explains the periodicity of the FRB or the variability from one cycle to another.
We can rule out relativistic spin precession binary model.
Lastly, we draw similarities between FRB~20180916B and a X-ray binary system, Her X~1, and explicitly note that both the sources exhibit a similar form of PA variability.
}

\begin{document}
%\linenumbers
%\label{firstpage}
%\pagerange{\pageref{firstpage}--\pageref{lastpage}}
\maketitle

%%%%%%%%%%%%%%%%%%%%%%%%%%%%%%%%%%%%%%%%%%%%%%%%%%
%%%%%%%%%%%%%%%%% BODY OF PAPER %%%%%%%%%%%%%%%%%%
\section{Introduction}

%% introduce R3, R3RM paper
\par Fast Radio Bursts (FRBs) are short duration transient events that occur in the radio regime of the electromagnetic spectrum \citep{lorimer2007,19PetroffReview,22PetroffReview,22BailesReview}.
\frb~\citep{chime2019b} is a repeating FRB source that emits bursts within a window which is (i) periodic with a period of 16.34 days \citep{20R3Period} and (ii) chromatic such that it shifts in time and varies in duration with observing frequency \citep{21PastorR3,21ZiggyR3,23BethapudiR3}.
Even with over 100 activity windows having been observed since its discovery, there has been no measurable change in period \citep{23SandRate,24LanPdot}.
%The source was localized to an edge of a spiral galaxy at a redshift of $0.0337$ with European Very Long Baseline Interferometry Network \citep[EVN;][]{20MarcoteR3}. 
%\citet{21TendulkarR3}, using Hubble Space Telescope (HST) observations of the host galaxy, reported the FRB location to be around sixty parsec away from the nearest star-forming region, and placed constraints on the age of the FRB source to be of the order of 10 Myr.
%\citet{22KaurR3}~performed HI spectroscopy observations using upgraded Giant Metrewave Radio Telescope (uGMRT) with which they report the host galaxy to be gas-rich and has a low Star Formation Rate (SFR), suggesting the host galaxy might have undergone a minor merger in recent past at the FRB location.
% HST constraint
% MWL constraint
Published bursts of \frb~across wide frequencies exhibit flat or unchanging Position Angle (PA) over burst timescales \citep{21NimmoR3,21PastorR3,21ZiggyR3,23BethapudiR3,23McKinvenR3,21SandR3,24GopinathR3}.
Moreover, the Rotation Measure (RM) of the FRB has been known to vary in a step-like fashion \citep{23McKinvenR3,24GopinathR3,24BethapudiRM,24NgRM}.
The causes for periodicity, chromaticity, lack of any variability in period, and step-like variability in RM is not known.
There have been many progenitor models to explain \frb\ \citep{20LevinPrecession,20ZanazziPrecession,20PazULPM,21WadaBinary}. 
A subset of those models explain periodicity using some form of dynamics such as rotation, precession, or binary motion.
We call such models collectively as dynamical models.
Each dynamical model has its own geometry and predicts unique temporal variability of PA on short and long timescales.
We take inspiration from \citet{21DongziPA} and study the temporal PA variations and compare what is observed and what is predicted by the various dynamical models.
This way, we hope to understand the geometry of the system and produce novel knowledge about the source.

% pa-longterm-pa-study
\par The PA of the polarized light measures its orientation with respect to the plane of the sky.
%Having sufficient linearly polarized signal content enables measuring the Position Angle (PA) of the emission, which tracks the orientation of the linear polarized signal. 
If the nature of emission is such that the PA is tied to the geometry of the system, then by studying the PA variations, we can track the geometric variations.
Such insights into geometry have already demonstrated by studying the PA variations of pulsars.
Pulsars are fast spinning neutron stars, which emit electromagnetic radiation, primarily in radio \citep{handbook}.
The radio emission is thought to originate from the open magnetic field lines of its dipolar magnetic field. 
The open fields lines are co-rotating with the spinning neutron star. Therefore, the PA of the emission shows the imprint of the co-rotation.
This imprint is modeled with the Rotating Vector Model \citep[hereafter RVM;][]{rvm1969,EverettWeisberg2001}.
%Given the orientation of the dipole field of the neutron star and viewing geometry, the PA sweep against phase of the rotation can be completely understood by RVM.
Recent works, such as \citet{23JohnstonTPARVM,24JohnstonRVM}, have shown that RVM can explain the observed PA sweep in over 80\% of the pulsars in the Thousand-Pulsar-Array program \citep{tpa}.
Moreover, long term studies of PAs of pulsars and magnetars \citep[highly magnetized neutron stars;][]{17KaspiMag} have elucidated novel geometrical phenomena. 
For instance, 
\citet{19Desvignes}~performed RVM modeling of PA data of a binary pulsar system PSR~J1906+0746 over a time span of 10 years and measured drift of the spin axis due to relativistic spin precession.
Similarly, \citet{24MengSP}~studied the relativistic spin precession of a double neutron star system PSR~J1946+2052 using PA data collected over three years. 
Additionally, by fitting RVMs to PA data of magnetar XTE~1810-197 at different MJDs, \citet{24DesvignesMag}~reported how the magnetar relaxes its internal structure over the course of several days after X-ray burst activity. 
%Morever, \citet{24DesvignesMag}~attribute the change to be due to free precession.
Clearly, long term modeling of PA data lets us study the geometric variations exhibited by the emitting body, provided the PA tracks the geometry. 
%The crux of this paper to perform long term modeling of PA data from a new class .

\par Features seen in PAs of FRBs could also be indicative of geometry, but connecting the PAs of the FRBs to a geometry has not been straightforward.
So far, RVM modeling has only been attempted with few non-repeating FRBs that display ``informative'' (non-flat) PA sweeps.
FRB~20221022A~is a non-repeating FRB detected by CHIME/FRB, which presented a S-curve like PA sweep similar to what is predicted by RVM \citep{24MckinvenPA}.
Furthermore, \citet{24MckinvenPA}~perform RVM modeling assuming different duty cycles and attempt to constraint the geometry.
In addition, \citet{24BeraTwin}~reports discovery of a non-repeating FRB~20210912A by ASKAP which exhibits PA variations between its two sub-components. \citet{24BeraTwin}~tried to explain the variation by fitting RVM to the PA time series.
However, such geometric origin need not always be the case. 
For example, \citet{24BeraLine} describe two non-repeating FRBs whose polarization features evolve within burst duration possibly indicating a plasma birefringent propagation medium.
In this case, the variation of PA need not exclusively be tracking the geometry.
However, such exotic polarization features and propagating media are uncommon in FRBs, and \frb~does not exhibit any of such features.
All in all, a long term study of PAs of repeating FRBs has not been yet attempted, which this paper tries to do with \frb.
In Sect.~\ref{sec:da}, we describe the collection of bursts and all the analysis steps we performed to accurately measure PA which can be compared between observations. 
We study the variability of PA across various time scales and measure the rates of the variability in Sect.~\ref{sec:pavar}.
We compare the observer PA variability with that which is predicted by the dynamical models in Sect.~\ref{sec:models}.
Lastly, we present our discussions and conclusions in Sect.~\ref{sec:dis} and Sect.~\ref{sec:con} respectively.

%% discuss FRB PAs
%% R1, R67, 
%\par Non-repeating and repeating FRBs alike display wide diversity of PA variability.
%% nrFRB flat and unflat
%\citet{24PandhiFrbPol}~studied the PA curves of a sample of non-repeating FRBs detected by CHIME/FRB and report that more than 80\% have flat PAs.
%From the DSA-110 non-repeating FRB sample, \citet{24ShermanFRB}~suggest significant PA variations only in a small fraction (12\%) of the sample.
%In case of repeating FRBs, \citet{18GajjarR1}~reported FRB~20121102A~possesses flat PAs for diverse morphologies of bursts.
%\citet{21HenningR67}~reported that each individual bursts of repeating FRB~20201124A had unchanging PA however the PA changed drastically from burst to burst, independent of time separation of bursts.
%However, only recently, \citet{24NiuR67}~reported for the first time orthogonal PA jump for that repeating FRB.
%\citet{20Luo0301} reported a non-flat PA sweep for FRB~20180301A.

% SB: strange that R117 Zhang paper themselves do not comment anything about PA evolution

%% that para merged with this

\section{Data and Analysis} \label{sec:da}
%% use calibrated bursts from 24BethapudiRM
%% convert the bursts from psrfits to timer
%% because of some weird bug
%% apply the calibration solution same as was done in 24BethapudiRM
%% Measure RM per observation
%% PA correction : to make it absolute
%% final PA

\par This section describes all the data processing steps taken to measure PAs of the bursts.
%We start from calibrated bursts presented in \citet[hereafter \citetalias{24BethapudiRM}]{24BethapudiRM} which have been detected using the uGMRT \citep{gmrt1991,ugmrt} in Band~4 (550 MHz to 750 MHz).
%We first account for inter-observation PA offsets, so that PAs measured in different observations which are conducted on different days, can be meaningfully studied. 
%Then, we test and show that all the bursts within an observation fit with one RM and one PA.
%This way, we hope to mitigate the degeneracy between RM and PA, on each observation, we employ two methods to robustly estimate one RM per observation and then measure PA using the estimated RM.
%Lastly, we produce two PA datasets one from each method used to estimate RM.
%\subsection{Bursts} \label{ssec:bursts}
%\par We use the bursts presented in \citet[hereafter \citetalias{24BethapudiRM}]{24BethapudiRM}. 
All the bursts were detected with the uGMRT \citep{gmrt1991,ugmrt} in Band~4 (550-750 MHz) with Phased Array mode of operation, the details of which can be found in \citet{24BethapudiRM} (hereafter \citetalias{24BethapudiRM}).
Table~1 of \citetalias{24BethapudiRM} lists all the observations and number of bursts per each observation.
It also lists various auxiliary sources, such as noise diode, quasars and known pulsars, which have been used in flux/polarization calibration and testing the performance of the search pipeline.
In this paper, we show the activity phase, the number of bursts and the maximum time separation among bursts per each observation in Table~\ref{tab:ppdata}.
We use the periodicity model presented in \citet{23SandRate} to compute Activity Cycle and Activity Phase of the bursts.
That is, we use reference MJD of 58369.4 and period of 16.34 days.
Activity Phase is a number between 0 and 1, which linearly maps to the period.
Activity Cycle is the integer number of periods elapsed since the reference MJD.
We first convert the bursts from \cmd{psrfits} to \cmd{timer} format using \cmd{pam} \citep{psrchive}.
Then, we apply the calibration solutions as derived in \citetalias{24BethapudiRM}~using \cmd{pac}, which also performs parallactic angle correction.
From hereon, we only refer to the calibrated burst archive in \cmd{timer} format.
%This way, we acquire not only the PA but also its error.
%Conversion to \cmd{timer} is necessary to accurately compute the parallactic angle corrections. because we noticed a bug in which observatory coordinates are not properly passed when using \cmd{psrfits} format which causes incorrect parallactic angle corrections.

\subsection{PA calibration} \label{ssec:paoffsets}

\par PAs are measured with respect to the sky frame.
To meaningfully compare PAs measured from different observations, PA measurements have to be in the same sky frame.
Therefore, we need to perform PA calibration so that PA measurements of a some known source with known and fixed PA from different observations are consistent and representative of the true PA value.
This is usually achieved using polarized quasars, such as 3C138 and 3C286, whose PAs are modeled and fitted for as a function of observing frequency \citep{PerleyButler2014Pol}.
However, we do not have multiple observations when 3C138 was observed. 
Therefore, we employ a different strategy here. 
We make use of known pulsars, PSR~B0329+54 and PSR~J0139+5814, to perform PA calibration.
That is, we make use of PA-sweeps of the pulsars to perform PA calibration.
Our choice of using pulsars and not standard quasars implies our PA measurements although consistent among themselves, are not in any absolute frame as defined in \citet{PerleyButler2014Pol}.
That is, we have only taken care of inter-observation PA offsets and there may be a global PA offset which would bring them into the absolute reference frame of \citet{PerleyButler2014Pol}.
Naturally, PA-calibration and thus further analysis is only done with those observations where we have observed either of the two pulsars.

\par As our PA references, we use European Pulsar Network Database (\cmd{epndb}) 610 MHz pulse profiles of PSR~B0329+54\footnote{\url{https://psrweb.jb.man.ac.uk/epndb/\#gl98/J0332+5434/gl98_610.epn}} and PSR~J0139+5814\footnote{\url{https://psrweb.jb.man.ac.uk/epndb/\#gl98/J0139+5814/gl98_610.epn}}. 
We fold all the B0329+54 and J0139+5814 scans \citepalias[see Table 1, auxiliary sources column][]{24BethapudiRM} using \cmd{dspsr} \citep{dspsr} and 
calibrate using the same procedure and calibration solution we used to calibrate the bursts from the respective observations.
We also use same $QU-$fitting procedure \citepalias{24BethapudiRM} to estimate RM and also correct for it using \cmd{pam}.
Then, we phase align all the pulsar archives with their corresponding \cmd{epndb} reference pulse profile. We do the phase alignment by minimizing the cross-correlation between Stokes-I integrated profiles.
Now, measuring position angle correction is simply a matter of measuring the offset in the position angle sweep of the observed pulsar with respect to the reference PA sweep. 

\par In the end, there can only be one PA reference. 
The \cmd{epndb} pulsar profiles of PSR~B0329+54 and PSR~J0139+5814 need not be in the same reference frame themselves and have to be brought to the same frame.
To do so, we note that we have four observations (MJD~59274, MJD~59275, MJD~59944, MJD~59993) where we have scans of both of the test pulsars. 
We measure PA corrections from each of the test pulsars from all the four observations, which are tabulated in Table~\ref{tab:pacorr}.
We note that the difference between the PA corrections found using PSR~B0329+54 and PSR~J0139+5814 is consistent with mean of $42.50$ and standard deviation of $1.75$ deg.
This consistency in the difference across multiple MJDs strongly convinces us that (i) PAs can be calibrated, and (ii) a single PA reference can be constructed using either of the test pulsars.
Therefore, we set \cmd{epndb} PSR~B0329+54 as our main reference and correct \cmd{epndb} PSR~J1039+5814 by $42.5$ deg so that it agrees with the PA frame of \cmd{epndb} PSR~B0329+54. 
Then, for all the observations where we use PSR~J0139+5814 to do PA calibration, we use this corrected \cmd{epndb} PSR~J0139+5814.
We show the phase-aligned Stokes-I and PA calibrated PA-sweeps for each of the pulsar scans from different observations in Fig.~\ref{fig:pacorr}.
The reference source and the corrections for each observation are shown in Source and Correction columns of Table~\ref{tab:ppdata}.

\par Lastly, we note that PSR~B0329+54 has a complicated PA sweep and also is seen to undergo mode changes within our dataset. Moreover, its peak linear polarization fraction is not high. 
However, we note that for all the observations where we use PSR~B0329+54 for PA calibration, the Stokes-I profile and the PA-sweep agree well (see Fig.~\ref{fig:pacorr}).
In hindsight, PSR~B0329+54 might not have been an ideal choice.
%, however, since a large fraction of the pulsar observations are of PSR~B0329+54, we are forced to use PSR~B0329+54 as the primary reference.
Nevertheless, we are confident in our consistent position angle measurements and leave to a future work on acquiring absolute position angles. 

%\subsection{Flat PA within a burst}
\subsection{Measuring RM} \label{ssec:mrm}

\par When light passes through an ionized medium with magnetic fields threading along the propagating direction, the PA of light undergoes rotation. This effect is known as Faraday rotation. 
It is mathematically understood as, 
\begin{equation} \label{eq:f1}
    \mathrm{PA}_\lambda = \mathrm{PA}_\infty + \mathrm{RM}\lambda^2,
\end{equation} 
where $\lambda$ is the wavelength of light, PA$_\infty$ is the PA measured at $\infty$ frequency (or also the initial PA), and $\mathrm{PA}_\lambda$ is the PA measured at wavelength $\lambda$ after the light has propagated through the medium with rotation measure RM.
%The measured PA usually depends on $\lambda$.
Within our observing setup, which covers from 550 MHz to 750 MHz with around 2048 channels, PA measured in each channel is different and linear in $\lambda^2$.
PA$_\infty$ is the true PA of emission from the source and only that can be used to study long term variability.
Any finite frequency would have undergone some Faraday rotation and would not truly be the PA as emitted by the source.
Therefore, using Eq.~\ref{eq:f1}, PA measurements at any finite frequency are brought to infinite frequency to recover $\mathrm{PA}_\infty$, which are then used to study long term PA variability.
Using PA at infinite frequency allows to account for the fact that RM of \frb~is varying on timescales of days \citep[\citetalias{24BethapudiRM}]{23McKinvenR3,24GopinathR3}.
However, we emphasize that our measurement of \painf~is only as good as our measurement of RM.
Any noise in our RM measurement introduces noise in our \painf~measurement. Therefore, this sub-section wholly focuses on accurately measuring RM using Eq.~\ref{eq:f1}.

\par PA variability against wavelength (or frequency) is not exclusive to Faraday rotation.
In our observing setup, polarization is expressed in circular basis, which has two basis elements: Right Circular Polarization (RCP) and Left Circular Polarization (LCP).
The phase of the cross-correlation of the two basis elements is degenerate with the PA of light.
Moreover, the phase of the cross-correlation is also a parameter of the polarization calibration model, known as \cmd{dphase} \citep[denoted with $\phi$;][]{Britton2000}.
In our calibration strategy \citepalias{24BethapudiRM}, we modeled \cmd{dphase} as a linear function over the observing band.
That is, $\phi=\psi_r + \pi\mathrm{D}_\mathrm{ns}f_\mathrm{GHz}$, where $f_\mathrm{GHz}$ is frequency in GHz, 
the slope, $\mathrm{D}_\mathrm{ns}$, is known as cable delay measured in nanoseconds, as it directly corresponds to the delay between RCP and LCP.
And, the bias, $\psi_\mathrm{r}$, behaves like twice the PA. The definitions used here are the same as \citetalias{24BethapudiRM}.
%Using multiple noise diode scans taken during different observations, 
Using observations in which multiple noise diode scans were taken, we note that the parameters of the linear model - D$_\mathrm{ns}$ and $\psi_r$ - are consistent with one another. See Fig.~\ref{fig:dpalin} and Sect.~\ref{ssec:dphasemodeling}.
%In addition, using the same collection of scans, we show  the scan to scan \cmd{dphase} variations in Fig.~\ref{fig:dpavar}.

\par Nevertheless, we note departures from the linear modeling of \cmd{dphase}. 
Fig.~\ref{fig:dpadev} shows the difference between the modeled \cmd{dphase} and observed \cmd{dphase} across the bandpass from the multiple noise diode scans of different MJDs. Deviations from the linear assumption are seen predominantly at the band edges, although are not limited to the edges. 
When the bursts are calibrated with linearly modeled \cmd{dphase}, the deviations from the linear model interfere with RM measurements.
This is particularly exacerbated by the band-limited emission of some of the bursts, because when the frequency span with significant non-linear \cmd{dphase} deviations overlaps with the burst emission, it would cause $QU-$fitting algorithm to under-/over-compensate the RM parameter to improve the fit.
As an illustration, we show the frequency spans and RM measurements of a subset of individual bursts detected on MJD~59243, where we had around forty bursts in total in Fig.~\ref{fig:rmfreq}.
We also plot the inverse variance weighted average of RM using all the bursts with a vertical red line.
We only selected bursts that occupy  half of the total band.
From the plot, we clearly see that bursts that occupy the upper edge of the band overestimate the average RM, while bursts that occupy the lower edge of the band underestimate the RM.
These spurious RM measurements, which are non-physical, are caused by \cmd{dphase} deviations from the linear model. 
Therefore, we do not use individual bursts' RM measurements in any way for further analysis.
However, we note that the average computed using all the bursts is less impacted by this effect, as the average also uses bursts with large bandwidths, but nevertheless, if a subset of individual RM measurements are affected by \cmd{dphase} deviations, any ensemble average would also be affected. 
As mentioned before, our PA measurement strongly depends on our RM measurements. 
Because we see nonphysical biases in our RM measurements, we do not use individual RM measurements or any average of individual RM measurements to correct for Faraday rotation and measure PA.
%Moreover, inherent variability in D$_\mathrm{ns}$ caused by the observatory also causes deviations in RM which are not real.
%See Appendix~B of \citetalias{24BethapudiRM} which reports the standard deviation of $\mathrm{D}_{ns}$ to be $0.1$ ns.
%We illustrate this effect by simulating three bursts with emissions exclusively in the (t) top-half, the (b) bottom-half, and also (f) full-band having the same RM but different cable delays and then measuring RMs from each of them.
%We illustrate the bias in measured RM due to incorrect D$_\mathrm{ns}$ using simulated bursts in Appendix~\ref{a:ssec:simdelay}.

%% the system is stable, the deviations are stable, fit for one RM 

\par While individual burst RM measurements would suffer from \cmd{dphase} deviations, simultaneously fitting one RM and one PA to all the bursts from an observation would diminish the impact of such deviations.
In addition, any PA variability which is not linear in $\lambda^2$ is not Faraday rotation and will be seen in the difference between the data and the fit. 
Of course, such a fit would not be possible if the source inherently exhibits RM and PA variability within an observation.
Therefore, we hypothesize and attempt to fit one RM and one PA to all the bursts from an observation and then repeat the procedure for all the observations. 
We extract normalized Stokes-$q,u$, that is Stokes-$Q,U$ divided by $\sqrt{Q^2+U^2}$, from every time-frequency bin with burst emission for all the calibrated bursts in an observation and perform $QU-$fitting using the normalized Stokes-$q,u$.
We follow the same $QU-$fitting procedure used in \citetalias{24BethapudiRM}, where fitting is done with respect to the center frequency.
We also employ \cmd{equad} to artificially inflate the errors as already used in \citetalias{24BethapudiRM}.
We only use normalized Stokes-$q,u$ here to avoid having to fit a linear polarization fraction ($L_p$) parameter for each burst, which would require us to fit for large number of parameters.
Although the fitting is done using $QU$-fitting, we show the fit outcomes as PA against $\lambda^2$ in Fig.~\ref{fig:fitonerm}, as both are equivalent.
%We fit and plot PA according to Eq.~\ref{eq:f1}, as shown in Fig.~\ref{fig:pal2}.
The top panels show the PA as a function of $\lambda^2$ with data in black dots and fitted model in red line.
The bottom panels show binned averaged PA residuals also against $\lambda^2$.
The residuals are first binned according to the original frequency resolution of bursts and averaged to clearly show the departures from the linear \cmd{dphase} model.
The top x-axis in both the panels shows observing frequency.
$\lambda_c$ is wavelength of the center frequency.
It is evident that (i) all the bursts from an observation fit with one RM and one PA, and (ii) there are structures in error that are indicative of unmodeled \cmd{dphase} variation over frequency.
See, for example, the averaged errors in upper edge of frequency bands of MJDs~59243, 59568 and 60303, or the averaged errors in lower edge of frequency bands of MJD~59993.
Lastly, we report the RM and the error in RM in Table~\ref{tab:ppdata}.

\begin{figure}
    \centering
    \includegraphics[width=0.45\textwidth,keepaspectratio]{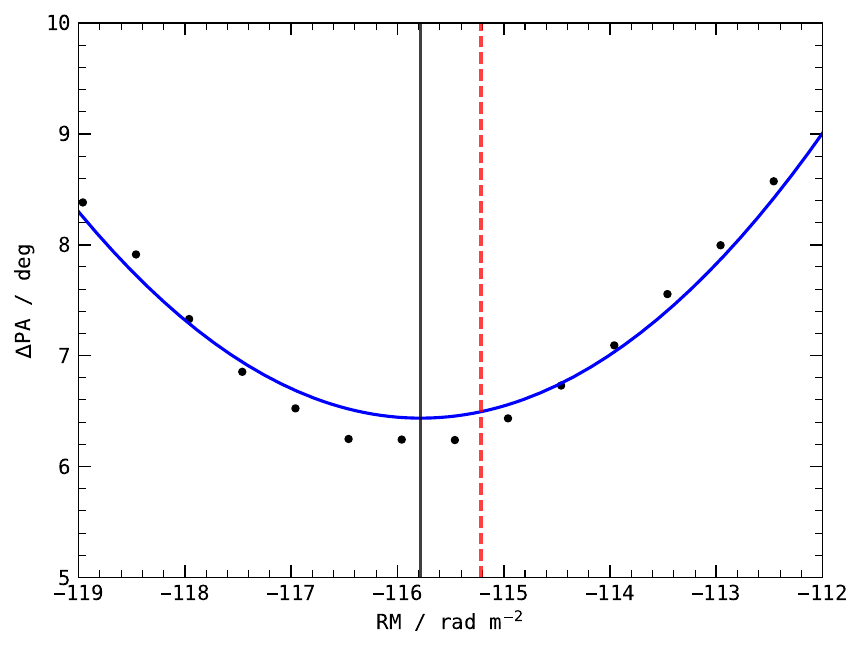}
    \caption{
    Applying one RM to all the bursts detected during the observation of MJD~59243 and measuring standard deviation of all the PA measurements from every timebin of every burst.
    Repeating this procedure for multiple RMs yields the black dots.
    The blue line is the fitted parabola. The vertical black line is the center of the fitted parabola. The red dashed vertical line is the RM measured after fitting using all the bursts. See Table~\ref{tab:ppdata} and Sect.~\ref{ssec:mrm}.
    }
    \label{fig:parabola}
\end{figure}

%% we do not know
\par We illustrate the difference between the above approach of fitting one RM to all the bursts and inverse variance weighted average in Fig.~\ref{fig:rmonewrm}.
The top panel shows the difference of one RM fitted to all bursts (denoted with \texttt{one}) and the weighted average (denoted with \texttt{wrm}) for all MJDs. 
The bottom panel shows the significance of difference. 
A horizontal black dotted line is drawn at $3\sigma$.
The number in parenthesis next to MJDs is the number of bursts detected.
The difference, although within $3\sigma$, significantly changes the measured PA. 
In our observing band, from 550 MHz to 750 MHz, an RM offset of even one unit causes an offset in \painf~by over 12 deg.
Therefore, we only use one fitted RM for all the bursts to measure PAs.
Moreover, we show the RM versus MJD using the measured RM, along with measurements from CHIME/FRB \citep{23McKinvenR3,24NgRM} and LOFAR \citep{24GopinathR3} in Fig.~\ref{fig:onermmjd}.

%% alternate evidence that one RM and one PA
\par Lastly, our choice of fitting one RM and one PA to all the bursts of an observation, which although works, has to be motivated from the data.
To this end, we perform an experiment using all the bursts taken on the observation of MJD~59243, where we have over forty bursts.
We apply one RM to all the calibrated bursts and measure the standard deviation of all the measured PAs from every time-sample of every burst. 
We repeat this procedure for different RMs.
When we plot the standard deviation, denoted as $\Delta$PA, against the applied RM, we get a parabola as shown in Fig.~\ref{fig:parabola} with black dots.
The blue line shows the best fit parabola and the vertical black line is the center of the parabola.
The vertical red dashed line is the fitted RM as described in the previous paragraph.
With this, we note that one RM exists which reduces the spread in PAs, suggesting that fitting one RM and one PA to all the bursts of an observation is valid.
Additionally, also see Fig.~\ref{fig:paspreadgame}, where by simulation, we show that spread in PA is only minimized if all the bursts indeed have the same RM and PA.
If any one of either RM or PA is not the same, we do not recover any parabola, which further indicates that our attempt is valid.

\subsection{Measuring PA} \label{ssec:mpa}

%\par Having measured one RM and one PA for each observation, and having shown that one RM and one PA fits for all bursts within an observation, we correct for the RM using \cmd{pam}.
%We extract PA and PA error time series using \cmd{pdv}. 
%The PA measured using \cmd{pdv} is with respect to the center frequency of the band, $650.048828125$ MHz.
%This PA is then brought to infinite frequency using Eq.~\ref{eq:f1}.
%All of commands used are part of \cmd{psrchive} software suite.
%% to epoch average
%We plot the individual burst PA time series consecutively with different colors for each observation in Fig.~\ref{fig:pashort}.
%We note, irrespective of time separation between the bursts, the \painf is unchanging. 
\par Having measured one RM and one PA for each observation, and having shown that one RM and one PA fits all the bursts within an observation, we bring the fitted PA from center frequency to infinite frequency using Eq.~\ref{eq:f1}.
The fitting, as mentioned before, is done with respect to center frequency of 650 MHz.
At the same time, we apply the PA calibration correction as determined in Sect.~\ref{ssec:paoffsets} to compensate for PA offsets.
The PA-calibrated, fitted PA at infinite frequency is denoted by \epainf.
The long-term analysis of \frb~is conducted with \epainf.
Alternatively, we can also extract using fitted RM and individual calibrated bursts.
We apply the fitted RM to all the bursts of an observation using \cmd{pam}, and extract PA and PA error time series using \cmd{pdv}
The PA measured using \cmd{pdv} is with respect to the center frequency of the band, which can be easily brought to infinite frequency using Eq.~\ref{eq:f1}.
Thereafter, we compensate for PA offsets.
The PA-calibrated, PA-measurements of individual bursts are denoted by \painf.

\par The average of \painf~is consistent with \epainf, that is, the average of \cmd{psrchive} derived measurements is consistent with fitted PA, therefore, we only use \epainf as a measure of average.
Instead of using the error in PA we get while fitting and propagating it, we use the standard deviation of \painf~as a measure of error in \epainf.
The reason is as follows: the standard deviation of PA measurements from individual bursts would include the spread in PA due to \cmd{dphase} variations, intrinsic RM variations as we fit one RM to all the bursts and any intrinsic PA variations.
For this same reason, we do not propagate the RM error into $\Delta$PA, as that would have already been considered when taking standard deviation.
Moreover, the fit error in PA is only of the order of 1 deg, which we think is heavily underestimated.
We tabulate the \epainf~and $\Delta$PA measurements in Table~\ref{tab:ppdata}.
%
%
%
%All of commands used are part of \cmd{psrchive} software suite.
%% to epoch average
%We note, irrespective of time separation between the bursts, the \painf is unchanging. 
%Finally, noting that one PA fits to all the bursts from an observation, we also calculate the average and standard deviation the \painf~of all the bursts in an observation and denote it by \epainf\ and $\Delta$PA, respectively.
\painf~measurements of each of the individual bursts are only presented in electronic form. See Data availablity.

\begin{table*}[h]
    \centering
%\resizebox{\linewidth}{!}{%
    \begin{tabular}{ccccccccccc} 
        \toprule \toprule
Phase & Cycle & MJD & Sep & N & RM & $\Delta$RM & \epainf & $\Delta$PA & Source & Correction \\ 
& & & hr & & rad m$^{-2}$ & rad m$^{-2}$ & deg & deg & & deg \\ \midrule
0.355 & 93 & 59894 & 1.3 & 4 & -61.54 & 0.23 & -51.93 & 3.86 & J0139+5814 & 49.39 \\
0.355 & 121 & 60352 & 0.0 & 1 & -56.48 & 0.50 & -10.49 & 4.94 & J0139+5814 & 78.47 \\
0.363 & 118 & 60303 & 2.1 & 11 & -58.79 & 0.53 & 1.85 & 7.48 & J0139+5814 & 69.94 \\
0.396 & 99 & 59993 & 1.9 & 3 & -52.63 & 0.36 & -21.08 & 4.22 & B0329+54 & 54.40 \\
0.397 & 55 & 59274 & 0.0 & 1 & -118.56 & 0.80 & -12.49 & 3.92 & B0329+54 & 61.42 \\
0.400 & 73 & 59568 & 0.7 & 10 & -68.97 & 0.48 & -26.08 & 5.27 & B0329+54 & 65.43 \\
0.403 & 96 & 59944 & 2.6 & 5 & -57.61 & 0.49 & -30.65 & 4.85 & B0329+54 & 47.88 \\
0.453 & 55 & 59275 & 1.3 & 6 & -113.04 & 0.39 & -80.98 & 4.98 & B0329+54 & 39.86 \\
0.494 & 53 & 59243 & 4.1 & 39 & -115.21 & 0.27 & -35.86 & 6.27 & B0329+54 & 65.93 \\
0.562 & 53 & 59244 & 0.3 & 4 & -113.64 & 0.56 & -65.21 & 6.11 & B0329+54 & 57.41 \\ \bottomrule
    \end{tabular}%
%}
    \caption{RM and \epainf~measured from every observation. $\Delta$PA refers to the standard deviation of the \painf~of all the bursts from an observation. 
    Cycle and Phase correspond to the Activity Cycle and Activity Phase of detection of bursts.
    Sep denotes the time separation between the first and the last burst of the observation. 
    Source and Correction refer to the PA correcting astrophysical source and the PA offset measured in degrees. N denotes the number of bursts detected in the observation.}
    \label{tab:ppdata}
\end{table*}

\subsection{Credibility of measurements} \label{ssec:credible}

\par We have had to use different calibration strategies using variety of different sources.
The bursts of MJD~59243, 59244, 59274, 59275 and 59568 were calibrated using a single scan of 3C138 taken at the beginning of the observation.
The bursts of MJD~59894, 59944, 60303, 60352 were calibrated using that noise diode scan that is nearest to when the burst occurred.
The noise diode failed during the observation of MJD~59993, therefore the bursts of that observation were calibrated using unpolarized quasar 3C48 and PSR~J0139+5814 pulsar, following procedure described in \citetalias{24BethapudiRM}.

%\par In addition to the deviations from linear model of \cmd{dphase} (Fig.~\ref{fig:dpadev}), from the multiple noise diode scans taken within one observation on different MJDs, we also observe variability in \cmd{dphase} data of one calibration scan to another (even with a separation of less than 30 minutes), which is not significant enough to change linear \cmd{dphase} model parameters.
\par From the multiple noise diode scans taken within one observation but on different MJDs, in addition to the deviations from linear model of \cmd{dphase} (Fig.~\ref{fig:dpadev}), we also observe variability in \cmd{dphase} data of one calibration scan to another (even with a separation of less than 30 minutes), although which is not significant enough to cause a change in linear \cmd{dphase} modeling and its linear parameters.
Fig.~\ref{fig:dpavar}~shows the variability as the difference between the \cmd{dphase} data of first scan and that of the consecutive scans for every observation.
While we only noted the deviations and variations in \cmd{dphase} in the case of multiple noise diode scans, as \cmd{dphase} variations are a property of the observing instrument, we expect our calibration solutions constructed from quasars and pulsars to also suffer from the same issue.
However, we reiterate that in spite of these variations, $\mathrm{D}_\mathrm{ns}$ and $\psi_r$ parameters are consistent (Fig.~\ref{fig:dpalin}) and do not show this variability reconfirming the validity of our calibration solutions.

\par In light of these instrumental deviations and variations of \cmd{dphase}, we investigate the impact this would have on our RM and PA measurements.
Narrow-band bursts show biases in RM measurements as noted in Fig.~\ref{fig:rmfreq}, however, broad-band bursts do not show those biases as \cmd{dphase} deviations do not mimic Faraday rotation since they are not linear in $\lambda^2$. 
We checked this by simulating bursts with known RM, PA and no structure, and superimposing the \cmd{dphase} deviations (as shown in Fig.~\ref{fig:dpadev}) into simulated bursts' PA as a function of frequency. Then, we recovered the true RM and PA, with differences in RM $\lesssim 0.2$ rad m$^{-2}$ and PA $\lesssim 0.4$ deg, from all the simulated bursts which reaffirms us that broad-band bursts are robust to \cmd{dphase} deviations.
Having shown that broad-band bursts are immune to \cmd{dphase} deviations, and since we fit for one RM and one PA using all the bursts of an observation, we are mimicking broad-band bursts.
To visualize this, we computed the cumulative distribution of the frequency spans of bursts detected within an observation in Fig.~\ref{fig:freqdist}. 
The black bold line shown the cumulative uniform distribution over the frequency band.
The collection of bursts from any observation occupies the whole band, and thus behaves like a broad-band burst, and must be immune to \cmd{dphase} deviations.
Therefore, we compare \epainf~across observations, but since our fitting procedure does not take into account \cmd{dphase} deviations, which would manifest as colored noise when $QU$-fitting, there might be some bias.
So, we err on the side of caution and only treat \epainf~variability across observations at this point as preliminary.
Accounting for the bias to robustly study the PA variability across observations would be taken up in a future work.

\section{PA variability} \label{sec:pavar}

\par We divide our discussion on PA variability in two parts: (i) intra-observation variability includes changes from burst to burst within a single observation (typically 2-4 hours) and (ii) inter-observation variability presents long-term changes on timescales $\geq$ day.

\subsection{Intra-observation} \label{ssec:short}
%% (1) flat within burst, flat among bursts, flat <= four hours 
%% (2) pa(time) RM flatness

\par
\begin{figure*}[h]
    \centering
    \includegraphics[width=0.95\textwidth,keepaspectratio]{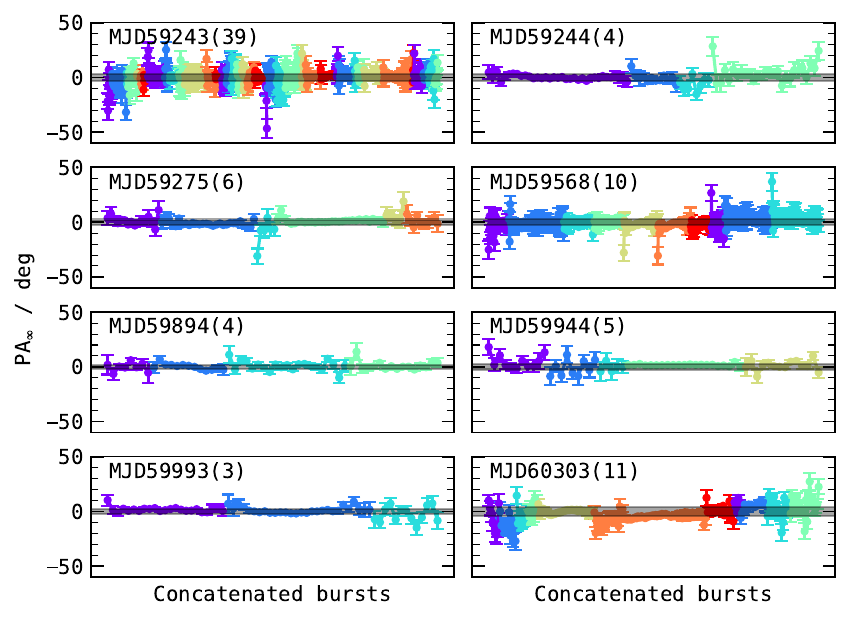}
    \caption{\painf~values of each burst plotted one after the another in individual subplots for every MJD. Each \painf~has been rotated by \epainf~taken from Table~\ref{tab:ppdata}. The different colors indicate different bursts. The shaded region around it is $\Delta$PA, as reported in Tab.~\ref{tab:ppdata}. The top-left text in each subplot is the corresponding MJD and the number of bursts in parenthesis.}
    \label{fig:pashort}
\end{figure*}

 In order to investigate burst-to-burst PA variations within an observation, we plot the \painf~time series of every burst consecutively with different colors in individual subplots for each observation in Fig.~\ref{fig:pashort}.
The \painf~time series are rotated by -\epainf from Table~\ref{tab:ppdata} so that they are all centered at 0. The gray shaded region is the $1\sigma$ error as presented in Table~\ref{tab:ppdata}. 
We note that \painf\ measurements are consistent with a single value even on the timescale of hours.
This is also evident from the fact that we were able to fit all the bursts with a single PA and RM value (see Sect.~\ref{ssec:mrm}).
That is, there is no sign of secular variability when the time separation between the bursts is of the order of hours (see `Sep' column in Table~\ref{tab:ppdata} for the time separations).
%By inspection, we report that \painf~does not on the timescale of hours. 
Since the maximum time separation in our dataset is about four hours (MJD~59243), we report an upper limit on variations of \painf to be $\leq 7$ deg within 4 hours ($\Delta$PA of MJD~59243, see Table~\ref{tab:ppdata}), and moreover, we note that this upper limit holds for all MJDs.

%\begin{figure}
%   \centering
%   \includegraphics[width=0.45\textwidth,keepaspectratio]{scifigures/paburstshort_59243.pdf}
%   \label{fig:pashort59243}
%    \caption{}
%\end{figure}

%% take closer look at MJD59243

\subsection{Inter-observation} \label{ssec:long}

%% (1) same cycle-different phase
%% (2) same phase-different cycle
%% (3) pa-phase difficult to interpret

%\subsection{Phase} \label{ssec:paphase}

\par We start with a reminder that all inter observation PA variability is preliminary at this point. We first study the PA variability directly over MJD and over Activity Phase.
PA variability could be tied with the periodicity of the FRB and must also be studied accordingly.
The periodicity allows us to break time into Activity Cycle and Activity Phase. 
PA can be independently varying over both Activity Phase and Activity Cycle.
To disentangle both notions of variability, we follow the logic of studying PA variability by keeping any one of Activity Phase or Activity Cycle constant and varying the other.
Therefore, in the following subsections we study the PA variations along three fronts: 
(i) the complete dataset versus MJD; 
(ii) the complete datasets, as well as measurements within a single Activity Cycle (intra-cycle), versus Activity Phase;
and lastly (iii) measurements taken around the same Activity Phase (inter-cycle). 
In all the above mentioned cases, \epainf~measured across multiple observations are compared. Which, as already mentioned in Sect.~\ref{ssec:credible}, might possibly carry some bias. Therefore, we caution the reader from overtly interpreting the results at this point.

\subsubsection{PA versus MJD} \label{ssec:mjdphase}

\begin{figure}[h]
    \centering
    \includegraphics[width=0.45\textwidth,keepaspectratio]{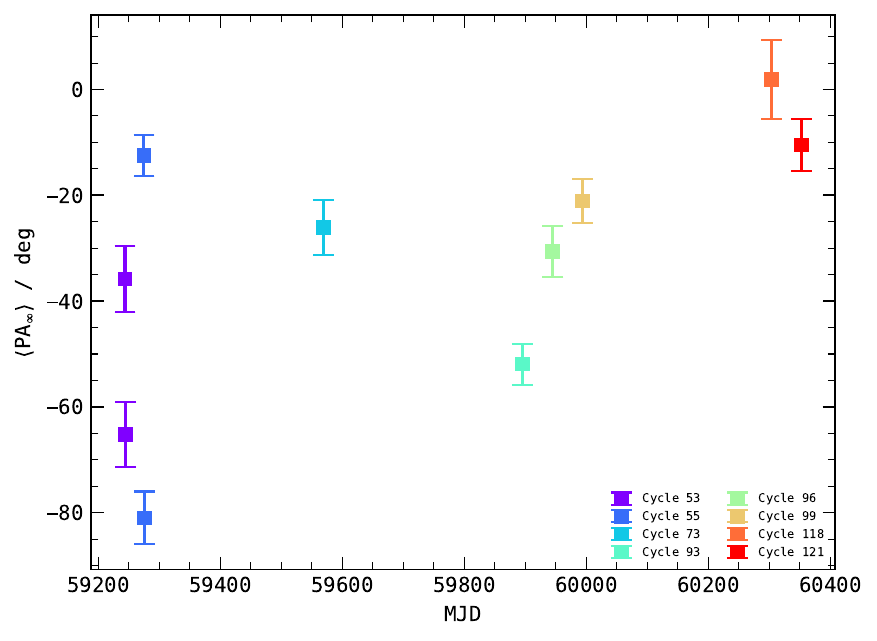}
    \caption{\epainf~versus MJD. The points are color coded according to their corresponding Activity Cycle.
    }
    \label{fig:pamjd}
\end{figure}

\par Fig.~\ref{fig:pamjd} shows \epainf~versus MJD. 
The colors indicate different Activity Cycles of the measurements, and the color-scheme used here is the same throughout the paper.
We emphasize again here that we have corrected for long term instrumental PA variation by aligning the PA swings of PSR~B0329+54 and PSR~J0139+5814 at different times (see Sect.~\ref{ssec:paoffsets} and Fig.~\ref{fig:pacorr}).
For \frb, we observe a total spread in PA of about 80 deg over the $\sim$1100 days of monitoring, but \epainf~versus MJD does not show clear evidence of a linear trend. However, we note that if there exists variability as a function of Activity Phase and/or Activity Cycle, looking at PA against MJD may hide clear trends of this variability.
%However, we note that when looking against MJD, we will have Activity Phase and Activity Cycle variability, if both exist, simultaneously which can hide clear trends of variability.

\subsubsection{PA versus Phase} \label{ssec:pp}

\par
\begin{figure}[h]
    \centering
    \includegraphics[width=0.45\textwidth,keepaspectratio]{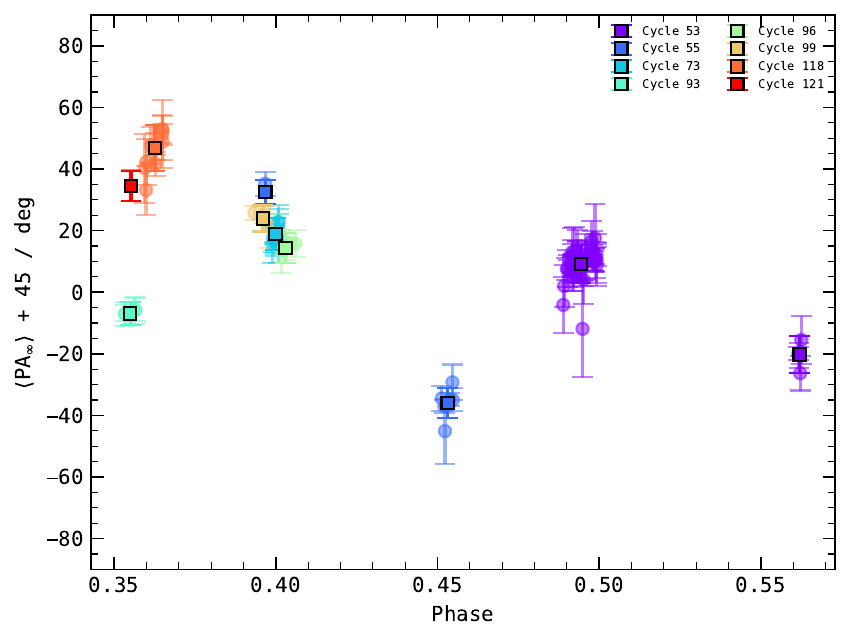}
    \caption{\epainf~versus Activity Phase of \frb~plotted with markers with black border. \painf~versus Activity Phase of the bursts plotted with faint markers.
    Both \painf~and \epainf~have been rotated by 45 deg.}
    \label{fig:pp}
\end{figure}

 We plot \epainf\ and \painf~versus Activity Phase in Fig.~\ref{fig:pp} as markers with black border and faint markers respectively.
Note that, to avoid PA wrap arounds, we have rotated \painf~and \epainf~by 45 deg.
The range of \epainf~measurements in both the datasets is around 80 deg.
Although, we have 10 measurements, they occur only at five distinct Activity Phases. 
%We note that the three clusters around phases 0.35, 0.4, and 0.5 are consistent with the same value, while the two clusters around 0.45 and 0.56 have a similar value. 

\par We have two instances with more than one measurement from the same Activity Cycle (see Table~\ref{tab:ppdata}). 
Two observations were made in (i) Cycle-53 on MJD~59243 and~59244 and (ii) Cycle-55 on MJD~59274 and~59275.
%The observation pairs in Cycle-53 and Cycle-55 are separated by 26.52 hours and 22.13 hours respectively.
%The phase spans of these pairs are also different. 
Cycle-53 observation pair phase ranges from $0.494$ to $0.562$ (26.52 hours), whereas Cycle-55 observation pair ranges from $0.397$ to $0.453$ (22.13 hours).
We study the variability within the same Activity Cycle, which we call intra-cycle variability, using the rate of change of PA.
In Cycles~53 and 55, the rates of change of PA computes to be $-58.8$ and $-29.2$ deg Activity Phase$^{-1}$ respectively, or in regular units, $-1.0$ and $-3.2$ deg hr$^{-1}$ respectively or 
For both the pairs, the significance of variability is at least $3\sigma$.
Interestingly, we note that rates of change of PA is different for these two pairs, which suggests that rate of intra-cycle variability depends on Activity Phase.

\par We test if intra-cycle variability is consistent with the observed non-variability of \painf $\leq 7$ deg over four hours of observation of MJD~59243 (Activity Cycle~53, see Sect.~\ref{ssec:short} for details).
%Refer to Sect.~\ref{ssec:short}.
With the intra-cycle variability slope of $-1.0$ deg hr$^{-1}$ from Cycle-53, in a span of four hours, we expect to see variations of around 4 deg. 
The $\Delta$PA of MJD~59243 observation (whose session was four hours in duration) is $6.27$ deg, which means observed intra-cycle variation is consistent at $1.5\sigma$ with our non-variability of PA observation.
This suggests that PA is varying against Activity Phase slowly enough that on timescales of four hours or less, the PA appears to be constant.

\subsubsection{Inter-cycle at a constant phase} \label{ssec:cc}

\par
\begin{figure}[h]
    \centering
    \includegraphics[width=0.45\textwidth,keepaspectratio]{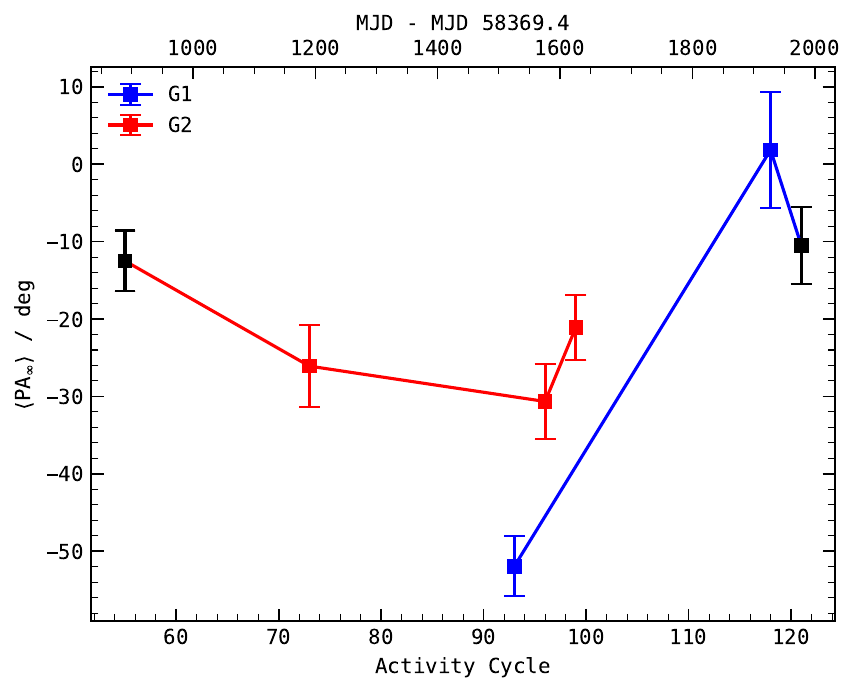}
    \caption{\epainf~versus Activity Cycle for the two phase clusters - \gcone~and \gctwo. \gcone~refers to the \epainf~measurements made at around Phase~0.359 (MJD 59894, 60303 and 60352) and \gctwo~at around Phase~0.400 (MJD 59274, 59568, 59944, 59993). See Sect.~\ref{ssec:long}.
    The black markers are used to dote \epainf~measurements with one burst. 
    The top x-axis is in MJD since the reference MJD of the periodicity model of \frb.
    }
    \label{fig:longtime}
\end{figure}

 Two of the five clusters in Activity Phase have more than one \epainf~measurement.
\epainf~measurements of MJD~59894,~60303, and~60352 are within a phase separation of $0.008$ (three hours), with a mean phase of 0.359. We denote this cluster by \gcone.
Also, \epainf~measurements of MJD~59274,~59568,~59944 and~59993 all are within a maximum phase separation of $0.006$ (2.75 hours) and a mean phase of 0.400, which we denote by \gctwo.
%The rest of the measurements are the only measurements at their respective phases.
%The rest of the measurements do not belong to any such phase cluster.
We plot the \epainf~versus Activity Cycle for both the clusters in Fig.~\ref{fig:longtime} in blue and red respectively.
The top x-axis is in MJD since the reference MJD of the periodicity model of \frb.
The \epainf~measurements where we only have one burst are shown with black markers.
We find that for both the clusters there is a clear sign of variability.
That is, \epainf~measurements taken at the same Activity Phase varies against Activity Cycle.
While we do not see a clear trend, if we ignore the \epainf~measurements with only one burst (shown with black markers), we still notice variability.
%However, those \epainf~measurements could very well be physical.
Therefore, we report inter-cycle variability and wait until more \epainf~measurements are made at similar Activity Phases to further characterize it.

\par Simply, as an order of magnitude estimate, we can compute the rates of change from the two clusters. 
The rates of change is around $\sim 0.1$ deg day$^{-1}$
%\par We can also compute the rates of change between every consecutive pairs of each of the clusters.
%For \gcone~with three measurements, the rate varies from $-0.2$ to $0.1$ deg day$^{-1}$.
%For \gctwo~with four measurements, the rate varies from $-0.05$ to $-0.007$ and finally to $0.1$ deg day$^{-1}$.
%The rates of variability are not constant between the two clusters, in addition also exhibit variability with time (Activity Cycle in this case).
For comparison, when studying \epainf~versus MJD, we noted 80 deg change over 1100 days, which amounts to a rate of change of $0.07$ deg day$^{-1}$ which is partly consistent with what we measure here.
The inter-cycle variability is of the order of deg day$^{-1}$, whereas the intra-cycle variability is of the order of deg hr$^{-1}$, which suggests there are two scales of variability.
Having said so, we do not have sufficient measurements to fully study the inter-cycle variability.
It could be simply PA variability linear in time which does not follow the periodicity of the source. 
At this point, we cannot comment further.
We again remind that we have performed PA calibration and have performed PA-calibration, that is accounted for any inter-observation PA offsets using PA-sweeps of pulsars (Sect.~\ref{ssec:paoffsets} and Fig.~\ref{fig:pacorr}).

\section{Constraints on progenitor models} 
\label{sec:models}

%\begin{table*}[h]
%   \centering
%\resizebox{\linewidth}{!}{%
%   \begin{tabular}{l|lcl} 
%        \toprule \toprule
%Models          &  & Assessment & Contradictory Evidence \\
%\hline
%Slow Rotation   &   & plausible & \\
%\hline
%\multirow{2}{*}{Free Precession} &magnetospheric emission&unlikely&lack of PA changes for bursts within hours$^{1}$\\
%\cline{2-4}
%&giant flares&unlikely&small Pdot$^{2}$\\
%\hline
%Forced Precessions& induced by discs/ companion& unlikely&active window drift against frequency$^{3}$\\
%\hline
%\multirow{3}{*}{Binary orbit} & \multirow{2}{*}{eclipse} & \multirow{2}{*}{unlikely} & widening of active window at low frequency\\
%&&&and lack of DM/RM change against phase$^{4}$\\
%\cline{2-4}
%&triggered emission& plausible &\\
%\bottomrule
%    \end{tabular}%
%    \caption{Summary of the current status of the models explaining the 16day periodicity. \\
%    \footnotesize{
    %$^{1}$:See Section 4.2;\\
%    $^{2}$:The expected Pdot is mentioned in, and the current published constraints are; \\
%    $^{3}$:The geometry only allow symmetric change of active window against frequency cite, which is against the observation cite;\\
%    $^{4}$ The plasma that introduces eclipse absorbs more at lower frequency and should introduce excess DM/RM. 
%    }
%    }
%    \end{table*}

\par In this section we connect the observed PA variability in conjunction with the periodicity of the \frb, to what is predicted by various progenitor models and place constraints on the geometry of the models. 
%Based on what we have observed, we propose to study the PA variability, and thus place constraints on progenitor models
Specifically, we test various progenitor models against the following observations: 
\begin{description}
    \item[\oone] Non-variability or limited variability of PAs $\leq7$ deg for up to at least four hours at all Activity Phases 
    \item[\otwo] Intra cycle variability with a rate of $\sim1$ deg hr$^{-1}$ or $80$ deg over $0.2$ Phase units
    \item[\othree] Tentative detection of inter cycle variability with a variable rate of $\sim0.1$ deg day$^{-1}$ which depends on Activity Phase
\end{description}
Since, we are treating all inter observation PA variability results as preliminary, we nevertheless discuss them in terms of models, but refrain from ruling out any model solely based on them at the moment (also see Sect.~\ref{ssec:credible}). 
%We again remind the reader about possible biases when interpreting inter observation PA variability (Sect.~\ref{ssec:credible}).
%since in addition to inter cycle variability, there is an \epainf~versus MJD variability which can only be explained if there is Cycle-cycle variability.

\begin{figure*}
    \centering
    \includegraphics[width=0.5\linewidth,keepaspectratio]{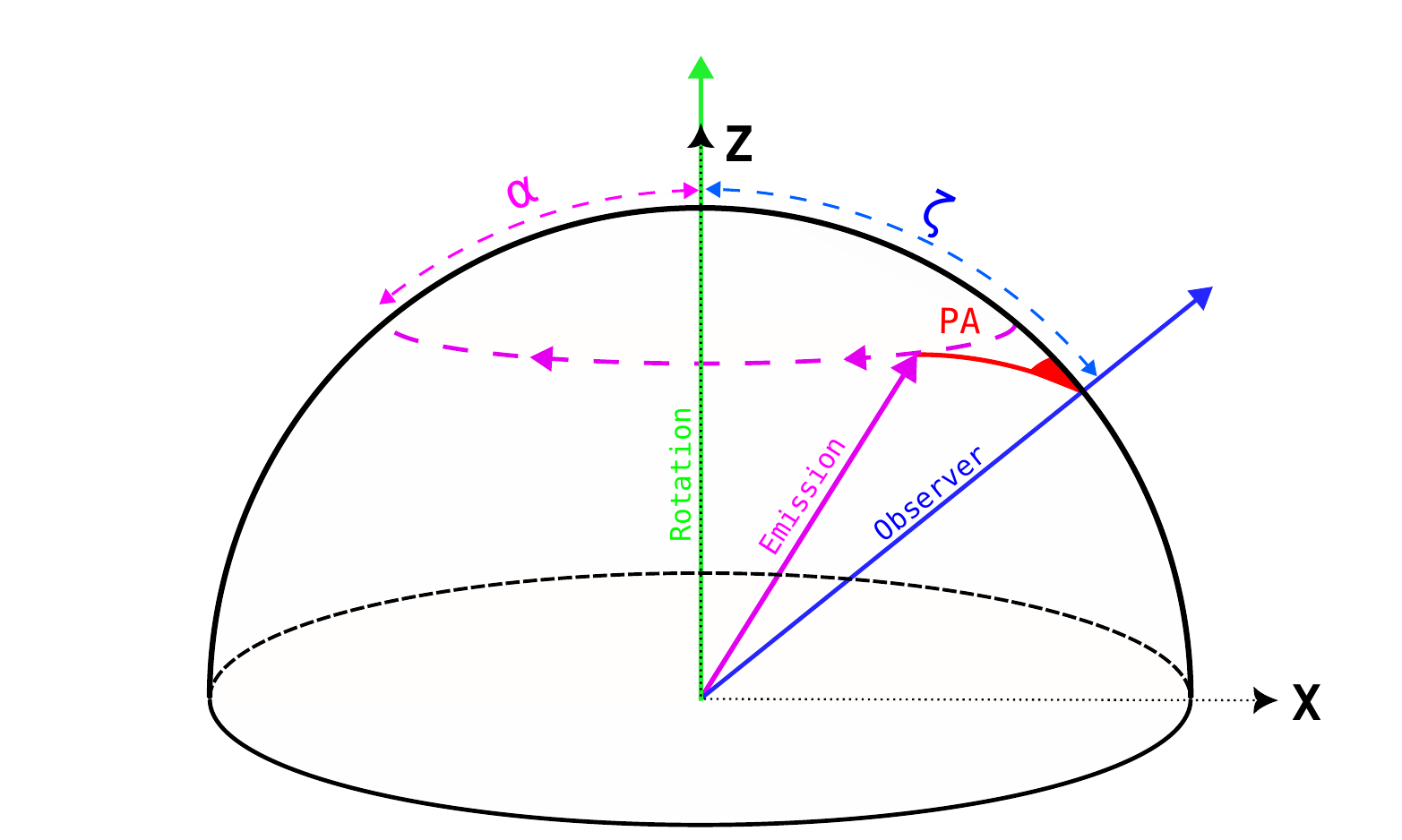}\hfill%
    \includegraphics[width=0.5\linewidth,keepaspectratio]{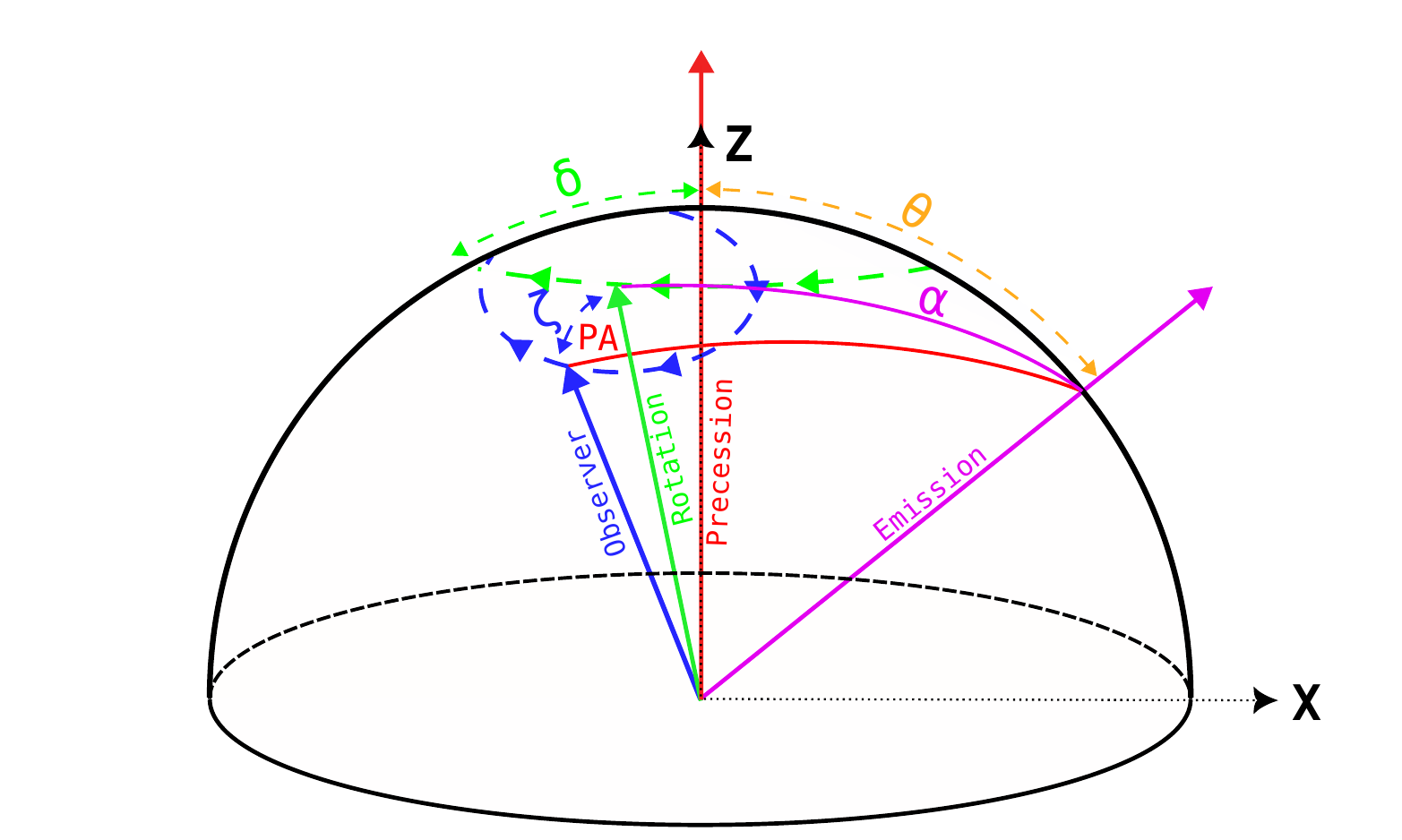}
    \caption{Cartoon illustration of rotation (\emph{left}) and precession (\emph{right}) geometries considered here. \emph{Left} is shown in the inertial coordinate system, which is fixed at the center of the neutron star. \emph{Right} is shown in the body fixed coordinate system.
    The X-axis is towards the right. The Z-axis is towards up.
    In case of rotation (\emph{left}), the rotation axis is along Z-axis. In case of precession (\emph{right}), the precession axis (symmetric axis) is along Z-axis.
    Motion is illustrated with large dashed line having arrows at regular intervals.
    Measured PA is appropriately marked. 
    }
    \label{fig:models}
\end{figure*}

\subsection{Slow rotation} \label{ssec:rot}

\par In a slowly rotating neutron star model for \frb, the rotational period is the period of the FRB \citep[$16.34$ days;][]{20PazULPM, 21DongziPA}. 
It is analogous to a typical pulsar except that the period is extremely large. 
We understand the geometry the following way:
an inertial frame is attached to the center of the neutron star such that the rotation axis is along the Z-axis (towards up), X-axis is towards right, and the observer is located somewhere in the XZ-plane.
We suppose that burst emission is directed along an arbitrary axis co-rotating with the neutron star, which we will call emission axis.
We illustrate the geometry in Fig.~\ref{fig:models} (\emph{left}).
The PA is the angle between the planes formed by observer and rotation axis, and observer and emission axis. It is illustrated in the same figure.
Moreover, the burst emission has the same PA along the emission axis at all times.
In case of pulsars, this emission axis is the dipolar magnetic axis and the emission mechanism is curvature radiation \citep{24MitraB0329}. 
As the neutron star rotates, the projection of the emission axis on the sky changes. 
This change in projection of the emission axis is tracked by the PA of the observed emission.
Note that this change in PA is not caused by the emission mechanism, but due to the rotation motion, hence is dynamical and geometrical in nature.
The projection is modeled with RVM \citep{EverettWeisberg2001}:
\begin{equation}
    \tan(\mathrm{PA}_0 - \mathrm{PA}) = \frac{ \sin(\alpha) \sin(\phi - \phi_0) }{% 
    \sin(\zeta)\cos(\alpha) - \cos(\zeta)\sin(\alpha)\cos(\phi - \phi_0)%
    }.
    \label{eq:rvm}
\end{equation}
$\alpha$ and $\phi_0$ are the magnetic axis co-latitude and longitude in the inertial frame respectively. $\zeta$ is the co-latitude of the observer, which is also known as viewing angle and $\mathrm{PA}_0$ is the projection of the spin-vector in the sky.
The magnetic field need not just be dipolar; axis-symmetric multi-pole magnetic fields also follow RVM \citep{23QiuMRVM}.
Therefore, our testing is still valid even if our assumption of dipolar magnetic field does not hold.
However, the same cannot be said for general multi-pole magnetic fields.

\par In this model, since the period is large, the PA would show a slow rate of change, which satisfies \oone.
In addition, we would expect intra-cycle PA variations, which satisfies \otwo.
Modeling of the intra-cycle PA variations with more measurements would help further constraint this model.
If the intra-cycle PA variations are RVM-like, the magnetic field configuration would be dipole-like.
%which are RVM-like, which could be what Fig.~\ref{fig:pp} is showing. 
However, in any case, there would be no variations in PA across Activity Cycle at the same Activity Phase, contrary to \othree.
In other words, according to this model, all the variations seen in \gcone~and \gctwo~are random.
This would be equivalent to the random variations in PAs of individual pulses of pulsars compared to that of the integrated pulse at the same phase \citep[See Fig. 2][]{24JohnstonRVM}.
Also see Fig. 1 of \citet{07MitraB0329} where the PA values show scatter about the RVM curve.
%, we cannot make conclusive statements as of now.
%We cannot yet perform any further testing as we have not sampled the active window enough. 
Lastly, we observe the range of \epainf~measurements span across 80 deg. 
Within the RVM model, the range of PA values is decided by $\alpha$.
Larger $\alpha$ leads to larger range of PA values.
However, with \epainf~measurements at only five distinct Activity Phases, any RVM modeling or constraining $\alpha$ would not be robust.
Therefore, we defer actual RVM fitting to future work.

\subsection{Fast precession} \label{ssec:fastp}

\par In the fast precession model for \frb, the burst emitting neutron star is rotating with a spin period and is also freely precessing with a precessional period \citep{20ZanazziPrecession,21DongziPA}.
Free precession implies there are no resultant torques on the compact object, which implies, the angular momentum is a constant of motion.
The precessional period is assumed to be the Activity Period of the FRB, and the spin period is predicted to be of the order of seconds \citep{20LevinPrecession,20ZanazziPrecession}.
Our geometry for precession is identical to that presented in \citet{21DongziPA}.
The geometry of this case is described as follows:
Instead of inertial frame, we attach a body fixed frame at the center of the neutron star such that the precession axis (or the symmetric axis, the axis along which the body is symmetric) is along Z-axis and the emission region (which is fixed in body fixed frame) is in XZ-plane.
The rotation axis is at a separation of $\delta$ \citep[$\theta$ in ][]{JonesPrecession}.
We make a reasonable assumption that rotation dominates the angular momentum, that is, the separation between the rotation axis and angular momentum axis \citep[$\hat{\theta}$ in][]{JonesPrecession}.
Therefore, PA is defined as the angle between the planes formed by observer and rotation axis, and observer and emission axis.
Because of our assumption, this PA definition is with respect to a constant reference axis (like angular momentum axis).
%This is guaranteed by our PA calibration step (see Sect.~\ref{ssec:paoffsets}).
%Now due to precession the rotation axis and angular momentum axis are no longer the same.
%The angular separation between the two is known as spin-misalignment angle, which we will denote by $\delta$ \citep{JonesPrecession,24HeylIXPE}. 
%In case of neutron stars, $\delta$ is expected to be an extremely small angle; for a typical ellipticity, $\epsilon$, of $10^{-6}$, $\delta \in \mathcal{O}(\epsilon^2)$.
%However, in the most general setting, it is treated to be a model parameter \citep{24HeylIXPE}.
$\delta$ is treated as a free parameter, which is known as the wobble angle \citep{JonesPrecession}.
We provide a cartoon illustration in Fig.~\ref{fig:models}(\emph{right}).
The observer in the body fixed frame executes the precession dynamics, which can be understood as the observer rotating about the rotation axis with rotational period and the rotation axis rotating about the precession axis with precessional period.
If we define the observer to be at $\zeta$ separation from rotation axis (like in Sect.~\ref{ssec:rot}), we note that co-latitude of observer as seen in the body fixed frame changes from $|\zeta-\delta|$ to $\zeta+\delta$ within a rotational period.
This co-latitude variability happens for every rotational period at every precessional phase.
Ultimately, it causes PA variations over the rotational period timescales.

\par We observed that PA is $\leq7$ deg on timescale of at least four hours (\oone) for multiple observations.
If the emission region was such that it only partially overlapped with the trajectory of observer, we would have been able to recover the periodicity.
Since, there has been no short timescale periodicity found in \frb, we suggest that the emission region is large enough to not cause any periodicity signature.
%The only way to reconcile this would be if the range of PA values is constrained to be within $\leq 7$ deg.
We also know that the modulation of PA values constrains $\alpha$, which gives us the constraint $\alpha\ll\zeta$. The maximum range of $\alpha$ is $|\delta+\theta|$, therefore, $|\delta+\theta|\ll\zeta$.
If variability of PA is constrained to be $\leq7$ deg (\oone), both $\delta,\theta\lesssim10$ deg, which causes $\zeta\gtrsim30$ deg.
In words, the rotation axis and emission axis are in close proximity to the precession axis, but the observer is far away from the rotation axis. 

\par Firstly, such a configuration cannot produce a modulation of 80 deg of \epainf~over Activity Phase (Fig.~\ref{fig:pp}).
In addition, a fast precession model also cannot explain inter cycle variation (\othree). 
The geometry of the entire system at any Activity Phase is identical for all Activity Cycles, which implies the observed PAs cannot be different for different Activity Cycles at the same Activity Phase.
The above inter-observation PA variability results directly make fast precession extremely unlikely. 
Regardless, even by only relying on intra-observation PA non-variability result (\oone), we can concretely rule out fast precession model.
With the above mentioned geometric constraint from \oone, it is almost impossible to provide an emission region which is large enough to cover all the spin-phases but which can only provide $30\%$ duty cycle in Activity Period.

\subsection{Slow precession} \label{ssec:slowp}

\par The slow precession model for \frb~is similar to the fast precession model, except that the spin period is the Activity Period of the FRB and the precessional period is much longer than the Activity Period.
Therefore, this model is the equivalent to a combination of slow rotation, which explains non-variability and intra-cycle variability (\oone~and \otwo), and free precession, invoked to explain inter-cycle variability (\othree).
However, precession would cause a gradual change in the Active window that has not been seen, which implies this model directly cannot explain the periodicity.
%Slow intra-cycle variability can be explained by slow rotation, as already done in Slow rotation case.
%Inter cycle variability is explained with the help of precession.
%the burst emitting neutron star is rotating with a spin period and is also freely precessing with a precessional period \citep{21DongziPA}.
%\par Slow phase-phase variability can be explained by slow rotation.
%Then, in order to explain cycle-cycle variations, we could have precession on top of slow rotation. 
%Moreover, in this case, the precessional period is longer than the total span of the dataset presented in this paper (1200 days), and we would predict that the inter-cycle PA variation would eventually show repetition. 

\par An alternate test for this model would be to study the inter-cycle variability. If that variability is indeed caused by precession, PAs at the same Activity Phase would exhibit sinusoidal variations with a period given by the precessional period, and the range of variation would be of the order of $\delta$ (defined in Sect.~\ref{ssec:fastp}).
We provide a simulation of this effect in Appendix~\ref{sec:spsim}.
We cannot yet estimate the precession period, however a lower limit of the precessional period would be the maximum time span of Activity Phase clusters - \gcone~and \gctwo, as we do not observe a full period yet, which is around $800$ days.
Now, if we assume an ellipticity of $10^{-4}$ \citep{20LevinPrecession} or $10^{-6}-10^{-8}$ \citep{24DesvignesMag}, and a rotational period of 16.34 days, the precessional period is $\sim\frac{10^{5}}{\cos(\delta)}-\frac{10^{10}}{\cos(\delta)}$ days.
Even if we take the smallest precessional period of $\sim 10^5$ days, a maximum time span of $800$ days would only mean a precessional phase span of 3 deg. 
For us to observe at least 50 deg of \epainf~variation within the 3 deg precessional phase span suggests an extremely large sinusoidal amplitude or $\delta$, which is highly unlikely.
Therefore, we conclude that maximum precessional period would be much less than $\sim10^5$ days.
We defer the attempt to accurately measure precessional period until inter cycle variability is measured with more significance and its variability is characterized.

\subsection{Binary models}
%% discussions
\par There are various binary models proposed to explain the periodicity and chromaticity of \frb~\citep{20IokaComb,21WadaBinary,BeXRB}.
In all such models, one of the two bodies is a compact object, and the bursts arrive only from this object. 
For instance, in \citet{20IokaComb} and \citet{21WadaBinary}, the emitting object is a neutron star, while in \citet{BeXRB}~the central object is a Be-type star with a disk and the bursts arrive from a compact object orbiting the star. 
In all cases, the period of the binary orbit is assumed to be the Activity Period of the FRB (16.34 days).

\par Whether or not PAs carry any information about the binary motion, depends on whether there is coupling between the emitting object's angular momentum and the system's orbital angular momentum \citep{1982GravPrece}.
In regular Keplerian dynamics, 
there is no such coupling. 
The orientations of the objects are not governed by their binary motion.
If PAs track the instantaneous orientation of the object, such PAs would not carry any information about the binary motion, and thus, cannot be used to probe binary models.
However, if we assume a binary model scenario where the PA of the emitted burst is purely a function of the true anomaly of the burst emitting object, we can perform some model testing. 
As the orbital period is the same as Activity Period, the predicted intra-cycle PA variability will match the observed variability.
Then to explain inter cycle variability, we will need to invoke some form of binary precession.
Modeling and testing such scenarios is a work in itself and requires much more measurements for any meaningful test. Hence, it will be taken up in future work.

\par That said, there are PA models for relativistic spin precession  \citep{KramerWex2009},~which have been used for Double Neutron Star systems such as PSR J1906+0746 \citep{19Desvignes} and PSR~J1946+2052 \citep{24MengSP}.
In this case, we suppose the orbital period to be Activity Period of the FRB.
The lack of any PA variations over rotational period timescales constrains $\alpha$, just like in the case of Fast precession (Sect.~\ref{ssec:fastp}).
In addition, due to relativistic spin precession, the projection of the spin vector ($\Psi_0)$ changes all the time throughout the orbit on spin-precession timescale \citep[Eq.~6;][]{24MengSP}.
That is, $\Psi_0$ is periodic with precessional period and does not depend on orbital phase (which is Activity Phase in this case).
However, we do not see any such signature in \epainf~versus MJD (Sect.~\ref{ssec:mjdphase}) and we know there are two scales of variability - intra cycle and inter cycle are tied with the periodicity of the FRB (orbital period).
Therefore, we do not suppose a relativistic spin precession model can explain the observed PA variations.

\section{Discussions} \label{sec:dis}

%However, even under this scenario, we cannot explain cycle-cycle variability.

\subsection{APERTIF PA sample}
\begin{figure}
    \centering
    \includegraphics[width=0.45\textwidth]{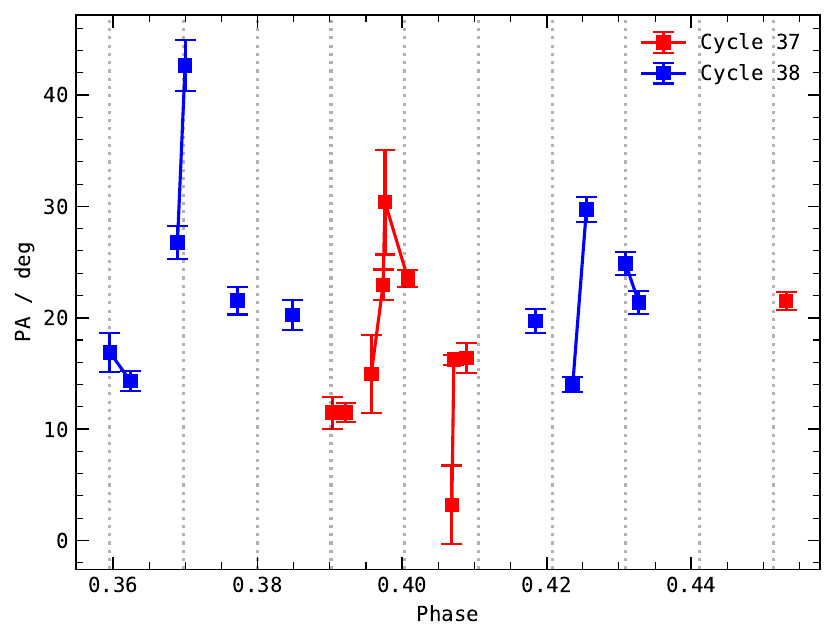}
    \caption{PA measurements of bursts detected by APERTIF at 1370 MHz in Activity Cycle-37 and -38 \citep{21PastorR3}. 
    Line connecting squares denotes measurements taken in the same observing session.
    The black dotted line is drawn for every four hours of phase-separation.}
    \label{fig:pm21}
\end{figure}

\par \citet{21PastorR3} measured PAs from the bursts detected using APERTIF at 1370 MHz at different Activity Phases and Cycles.
They report a range of PA values of around $44$ deg over a phase span of 0.1 (1.5 days), while we report a range of \epainf~values of around 80 deg (Fig.~\ref{fig:pp} and Sect.~\ref{ssec:pp}) and a phase span of 0.2 (3.25 days).
Moreover, \citet{21PastorR3}~have multiple PA measurements within Cycle-37 and -38, which we use here to test intra-cycle variability.
We plot the PA measurements of Cycle-37 (in red) and Cycle-38 (in blue) against Activity Phase in Fig.~\ref{fig:pm21}.
We connect those measurements that are done in the same observing session with a line.
Moreover, we add vertical dotted lines at intervals of 4 hours starting from the smallest phase.

\par Unlike in our analysis, \citet{21PastorR3} used a singular RM value for all the bursts, which were detected across multiple observations and cycles, and in addition, the bursts have been absolutely calibrated, i.e. PAs can be studied across Phase and Cycle.
While we observe PA variability $<7$ deg over four hour timescale, \citet{21PastorR3}~observes much rapid variability on these timescales. 
Furthermore, PAs measured from bursts within the same observing epoch also exhibit large variability, contrary to what we report.
We cannot test inter cycle variability as the \citet{21PastorR3}~dataset does not multiple measurements at the same Activity Phase.

%% how to reconcile?
\par This inconsistency can be explained in the following way. 
The PA measurements of \citet{21PastorR3}~have been made after correcting all the bursts with the same RM of $-115$ \urm.
All the measurements were made when the RM was only stochastically varying \citep{23McKinvenR3}. 
The standard deviation of the RM measurements in the stochastic regime (before MJD~59200) is about 3 \urm \citep{23McKinvenR3}.
PA measurements are at frequency of 1370 MHz, which even for a RM variability of the order of 3\urm causes a PA variability of about 8 deg.
Therefore, the observed variability by \citet{21PastorR3}~could be due to the use of a single RM for all observations rather than observation-specific RMs as used in this analysis.

\subsection{Her X-1} \label{ssec:herx1}
%% MWL 
%% MWL X-ray

\par Her X-1 is an intermediate mass accreting X-ray binary pulsar system in which the neutron star spins with a period of 1.24 seconds, the orbital period is 1.70 days and the whole system also shows a superorbital period of 35 days. 
Recently, by applying the RVM at different times to the neutron star's polarization profile, \citet{24HeylIXPE}~uncovered complex rotational dynamics involving rotation and precession of the neutron star, accretion disk and the companion, ultimately explaining the superorbital periodicity. 
%We would like to draw several parallels between Her X-1 and \frb.
%Firstly, we highlight that the ellipticity in the neutron star to be around $10^{-7}$.
The activity of Her X-1 switches states of activity multiple times over the superorbital periodicity \citep{Herx135day}.
It has two ON-states known in literature as main-on and short-on.
The activity variability of Her X-1 is caused by the crustal precession, which brings the active emission region into the line-of-sight of the observer at specific precessional phase \citep{24HeylIXPE}.
The extent of active region along the line-of-sight determines the scale and length of activity seen.
We can immediately draw many parallels between Her X-1 and \frb.
Firstly, the rate variability over superorbital period is analogous to rate variability seen over active window \citep[][\citetalias{24BethapudiRM}]{21PastorR3}.
Moreover, if the observer sees different emission regions at different phases, and if such emission regions emit at different frequencies, we could directly also explain chromaticity \citep[c.f.][]{21DongziPA}.

\par The superorbital periodicity of Her X-1 is known to be stable \citep{Herx1Pdot}. 
If the cause of that periodicity is indeed precession, one would assume various damping mechanisms to be at play, which would lengthen the period with time \citep{KatzPdot}.
However, \citet{24HeylIXPE}~show how torque interactions between disk and the star balance out the opposing torques such that the resultant motion is effectively torque free.
Now compare with \frb~for which \citet{23SandRate} and \citet{24LanPdot}~report that the period of \frb~is stable over at least two years.

\par \citet{24HeylIXPE} notes a change of about 8 deg in the $\Psi_0$ between two main-on states, which are separated by over an year \citep[Table~1, $\chi_p$ column of][]{24HeylIXPE}.
The PA variability gives a slope of $0.02$ deg day$^{-1}$.
This is largely consistent with the inter-cycle variability slopes we measure in Sect.~\ref{ssec:long}.
\citet{24HeylIXPE}~suggest that this variability is caused by a change in the angular momentum. 
While it is unclear at the moment what could be causing the angular momentum changes, we note that in free-precession regime the angular momentum is a constant of motion, as there are no torques involved.
As noted in the previous paragraph, even if there are multiple torques at play and all the torques balance out, the resulting system can maintain its period and also change its angular momentum.

\par Her X-1 is observed in optical and X-rays, with only marginal detection in radio at 9 GHz\citep{Herx1radio}. If \frb~is a source similar to Her X-1, we should expect to see temporal variability of optical and X-ray light curves over rotational and/or binary period timescales. 
This temporal variability should be persistent irrespective of the state of \frb~in radio.
However, no such optical or X-ray emission is detected. One reason could be the burst emission is beamed elsewhere. 
Although more likely, it could be due to large distance of \frb~(140 Mpc) compared with that of Her X-1 (7 kpc).
If we assume a X-ray luminosity of $\sim10^{37}$ erg s$^{-1}$ for Her X-1\citep{Herx1lumX}, and if we place Her X-1-like source at the distance of \frb, the observed X-ray luminosity is ten times less than the upper limit provided by Swift-BAT \citep{22LahaXray}.
Which suggests that large distance is hindering any X-ray detection.
And if Her X-1 \citep[or other such systems that exhibit superorbital periodicity, see][]{OtherSup} are similar to FRB-sources, we should expect to detect FRB-like bursts.
In so far, no such bursts have been detected, which could suppose that comparison is not so direct.

\subsection{Other repeating FRBs}

The current population of known repeating FRBs shows a diversity of PPA behavior within single bursts, as well as between bursts within a single observation. (We reiterate that this is the first robust study of PPA variation between different observations.) \rAO\ also exhibits flat PPA curves within individual bursts \citep{21HenningR1}, and notably, bursts within a single observation can also be fit with a single RM and PPA value \citep{18MichilliRM}. A small sample of bursts from \RGC\ show flat PPAs across a burst but jumps in the PPA value from burst-to-burst \citep{21NimmoM81}. This is in contrast to \rfast, \rss, FRB~2022012A, which show swings in PAs across the duration for some bursts and burst-to-burst variations within an observation \citep[e.g.][]{20Luo0301, 22JiangR67, 23ZhangR117}. That said, there is evidence that some of this may due to propagation effects through the plasma \citep{24Uttarkar,22XuR67}, which complicates the connection between the PPA and the geometry at the emission site. 

\frb's PPA properties are most similar to \rAO. Interestingly \rAO\ is also the only other repeating FRB with a known activity periodicity \citep{20RajwadeR1,25Braga}. With a sample size of two, it is difficult to judge whether this is simply a coincidence. Moreover, the host galaxies and local environments of the two sources are different. \frb\ is on the edge of a star forming region within a spiral galaxy and has no associated PRS \citep{20MarcoteR3}, while \rAO\ is in a star forming region in a low metallically dwarf galaxy and has a PRS \citep{17MarcoteR1,17TendulkarR1}. If they have the same progenitor, it needs to be able to be present in both these environments. 

\subsection{Future strategies}

%% how to test future strategies
\par Future observing campaigns should focus on studying intra cycle variability and robustly investigating inter cycle variability.
Studying intra cycle variability requires repeated observations of the source within the same Activity Cycle for multiple cycles.
Investigating inter cycle variability would require us to perform observations at the same Activity Phase on many different Activity Cycles.
Therefore, in addition to be able to track the source over long times, we would also need a sufficient burst rate to robustly measure the RM and track the PA variability. 
Given these requirements, we argue that the uGMRT is the first choice for this endeavor.
uGMRT can observe the source for around 11 hours on average, which means, it would require five days to cover the entire 2.1 days active window at 600 MHz.
In addition, uGMRT at 650 MHz observers a rate of well over one per hour \citepalias{24BethapudiRM}. 

\par Performing a long term PA variability study requires a stable PA reference frame so that we do not introduce any PA offsets.
In this study, we test for this stability by using multiple pulsar observations taken on four different epochs.
We find agreement between PA offset found using each of the pulsar PA sweep when compared to respective PA reference sweep (see Table~\ref{tab:pacorr}).
This agreement should be valid throughout the span of the study.
Considering that the reference PA frame used here is non-standard, it would be worthwhile to connect the reference PA frame with absolute reference frames such as those defined in \citet{PerleyButler2014Pol}.
Such connection would also be useful as an independent probe of PA validity.

\par We have to improve our calibration strategy.
From this study, we find that deviations from our linear model of \cmd{dphase} exist and impact our RM measurements.
Therefore, sophisticated calibration strategies could be first developed.
Alternatively, we can improve our fitting strategy to include Gaussian Processes that can can account for \cmd{dphase} variations to measure unbiased RM and PA estimates.
In addition, our procedure of fitting one RM and PA to all the bursts is althought robust, it requires sufficient number of bursts to perform as expected.
Future observing strategies could take that into account and observe the source for sufficiently long duration.
All in all, this work seeks to prompt many future work in this direction, and in particular, study such PA variations of other repeating FRBs. 
Of all the known repeating FRBs, in particular, we want to highlight FRB~20200120E which has been localized to a globular cluster \citep{KirstenM81}.
The properties of the bursts strongly suggest that young pulsar progenitor model produces bursts \citep{21NimmoM81}. If that is the case, PAs of the bursts would already carry a RVM-like signature, which can provide further evidence.

\section{Conclusions} \label{sec:con}

\begin{itemize}
%% data presented
\item We present a consistent, almost absolute calibrated  PA measurements of 84 bursts of \frb~collected over a time span of 1200 days with the uGMRT at 650 MHz.
We have performed PA calibration (Sect.~\ref{ssec:paoffsets} and Fig.~\ref{fig:pacorr}), 
showed and fitted one RM and one PA to all the bursts from an observation (Sect.~\ref{ssec:mrm}), and measured observation averaged PA and its error from all the observations (Sect.~\ref{ssec:mpa}).

%% slopes
\item We suggest that PA variability is best understood when studied in conjunction with the periodicity of the FRB. While we report the following findings, we remind the reader about possible biases in \epainf~which makes comparing \epainf~across different observations only preliminary at this point (Sect.~\ref{ssec:credible}):
\begin{itemize}
    \item The variability in PA is constrained to be $\leq 7$ deg over timescales of four hours (\oone) in all observations.
    \item The PA varies over Activity Phase, which we have termed as intra-cycle variability, with a variability of $\sim \pm 1$ hr$^{-1}$ (\otwo) or 80 deg over 0.2 Phase units.
    \item We also report that the PA varies with Activity Cycle, which we have studied by using \epainf~measured at the similar Activity Phases. We call this variability inter-cycle variability, which varies with time is of the order of $0.1$ deg day$^{-1}$ (\othree).
\end{itemize}

%% constraints on the models 
\item We are able to place constraints on various dynamical models for the FRB using the above mentioned findings. Given that our inter-observation PA variability is only preliminary, we test the models based on \otwo~and \othree~but do not rule out any solely based on them.
\begin{itemize}
    \item Slowly rotating model due to their slow rotation can explain limited variability over timescales of hours and intra-cycle variability (\oone~and \otwo).
    We conclude that slowly rotating model can explain \frb.
    However, such a model cannot explain inter-cycle variability (\othree) without suggesting the observed inter-cycle variability is purely by random variations.
    \item Fast precession models expect rotational-timescale PA variability, which is at odds with non-variability of PA on timescales of hours (\oone).
    Moreover, satisfying \oone~causes the model to contradict (i) lack of short time scale periodicity, (ii) $30\%$ duty cycle, and (iii) observed PA versus Activity Phase modulation (\otwo).
    In addition, fast precession model cannot explain \othree.
    
    Therefore, at this point, we successfully rule out fast precession model.
    %solely based on discord between \oone~and \otwo, we can successfully rule out this model.
    \item Slow precession model is equivalent to the combination of slow rotation and free precession, which can explain \oone, \otwo~and \othree. However, under this model, the Active window should change with time which has not been observed.
    Notwithstanding, this model predicts sinusoidal inter-cycle PA variations which can also be used to test this model with future measurements.
    \item Testing Binary models, where the binary orbital period causes the periodicity of the source, requires models to make PA predictions which as of now, none of the proposed models do. Testing such models will be done in a future work.
    Nevertheless, if we suppose \frb~is like a Double Neutron Star system exhibiting relativistic spin precession \citep{KramerWex2009}, \oone~can constrain the RVM geometry.
    However, \epainf~variations would not be following the periodicity of the FRB, which contradicts \otwo~and \othree.
\end{itemize}

Lastly, we compared \frb~with Her X-1 system and highlighted a lot of similarities in Sect.~\ref{ssec:herx1}. 
The most insightful of them all is the consistency between the rate of PA variability of Her X-1 measured in X-ray using IXPE \citep{24HeylIXPE} and the inter-cycle variability of \frb~measured in radio using uGMRT.
Continuing the long term study of PAs of both the systems can possibly provide us with new avenues to compare and understand both of them.

\end{itemize}

\begin{acknowledgements}
% SB thanks Joeri van Leeuwen.

We thank the staff of the GMRT that made these observations possible. GMRT is run by the National Centre for Radio Astrophysics of the Tata Institute of Fundamental Research.
VRM gratefully acknowledges the Department of Atomic Energy, Government of India, for its assistance under project No. 12-R\&D-TFR-5.02-0700. LGS is a Lise Meitner Independent Max Planck research group leader and acknowledges support from the Max Planck Society.
%%Part of this research was carried out at the Jet Propulsion Laboratory, California Institute of Technology, under a contract with the National Aeronautics and Space Administration.

\end{acknowledgements}

\section{Code availability}
All the code for data reduction and analysis, and to reproduce the results and plots and tables are shared with \hyperlink{zenodo}{zenodo}.

\section{Data availability}
The data underlying this article shared on any request to the corresponding authors.

\bibliographystyle{aa}
%\bibliography{frb20201124a}
\bibliography{cal,rest,frb}

\appendix

\section{Calibration verification} 

\subsection{PA calibration}\label{apex:paver}

\begin{table}
    \centering
    \begin{tabular}{cccc} \toprule \toprule
\multirow{2}{*}{MJD}& \multicolumn{3}{c}{Correction / deg} \\ 
 & B0329+54 & J0139+5814 & Difference \\ \midrule
%MJD & B0329+54 & J0139+5814 & Difference \\ 
%& deg & deg & deg \\ \midrule
59274 & 61.42 & 20.81 & 40.61\\
59275 & 39.86 & -1.75 & 41.62\\
59944 & 47.88 & 2.76 & 45.13\\
59993 & 54.40 & 10.78 & 43.62\\ \bottomrule
    \end{tabular}
    \caption{PA offsets determined from PSR~B0329+54 and PSR~J0139+5814 pulsars scans taken during the same observation. Difference column corresponds to the difference between the PA corrections derived from PSR~B0329+54 and PSR~J0139+5814.}
    \label{tab:pacorr}
\end{table}

\begin{figure}
    \centering
    \includegraphics[width=0.45\textwidth,keepaspectratio]{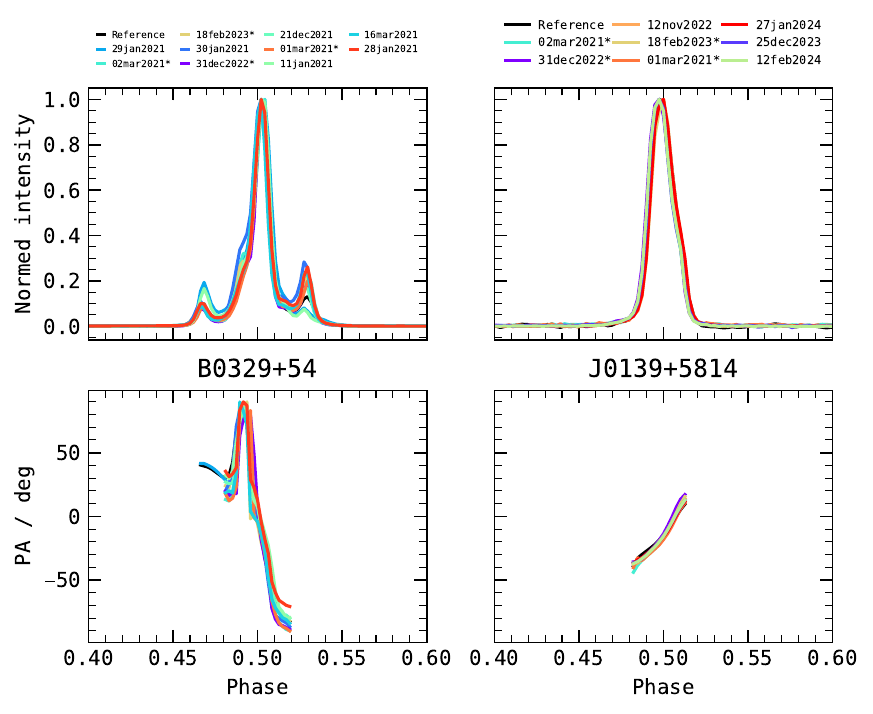}
    \caption{Stokes-I integrated frequency averaged pulse profiles (\emph{top}) and PAs against phase (\emph{bottom}) of PSR~B0329+54 and PSR~J0139+5814 observed at Band~4. See text for details.}
    \label{fig:pacorr}
\end{figure}

\par We plot the phase aligned Stokes-I integrated pulse profiles pulsars PSR~B0329+54 and PSR~J0139+5814 in the \emph{top} row of Fig.~\ref{fig:pacorr}. 
Each line is color-coded by the day of observation.
Reference refers to the \cmd{epndb} 610 MHz profiles.
The \emph{bottom} panel shows the matched PA against phase curves.
We verify our PA correction and measurement by noticing that when we transfer PA correction from PSR~B0329+54 to PSR~J0139+5814 on observations \texttt{01mar2021}, \texttt{02mar2021}, \texttt{12nov2022}, \texttt{31dec2022} and \texttt{18feb2023}, all the PSR~J0139+5814 PA phase curves agree.
Such an agreement is not expected and strongly convinces us that we have taken care of any intra-observation PA rotations.

%\section{RM-PA degeneracy}

\subsection{\cmd{dphase} modeling}
\label{ssec:dphasemodeling}

\begin{figure}
    \centering
    \includegraphics[width=0.45\textwidth,keepaspectratio]{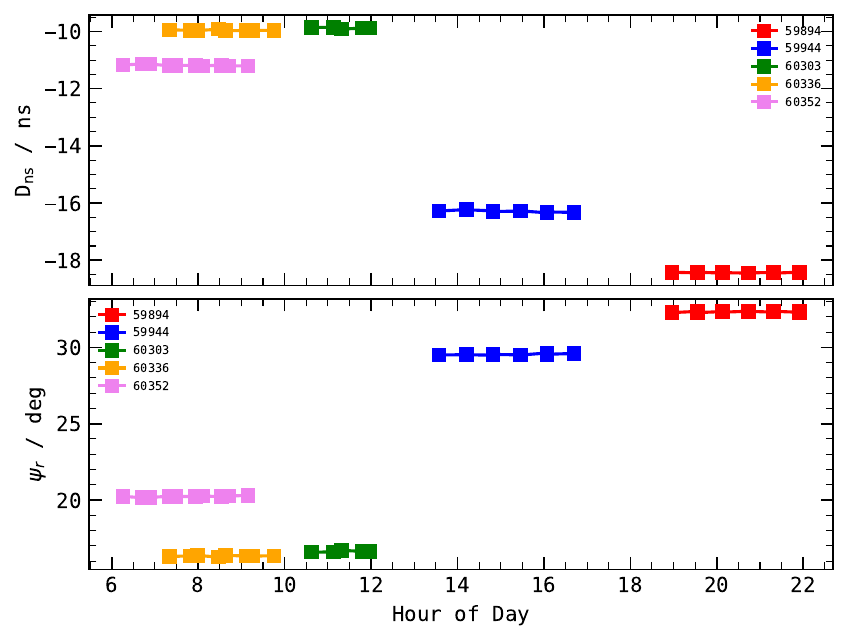}
    \caption{$\mathrm{D}_\mathrm{ns}$ and $\psi_r$ computed using noise diode scans from different MJDs plotted against Hour of Day when the scan was taken. The sign of $\mathrm{D}_\mathrm{ns}$ is purely by convention. }
    \label{fig:dpalin}
\end{figure}

\par uGMRT, being an interferometer, requires phasing of the antennas at regular intervals to produce and maintain a coherent beam on the source.
Here, we check if our \cmd{dphase} linear modeling is valid from scan to scan within an observation.
Fig.~\ref{fig:dpalin} shows $\psi_\mathrm{r}$ and $\mathrm{D}_\mathrm{ns}$ measurements from multiple noise diode scans taken during multiple observations.
The consistency not just in $\mathrm{D}_\mathrm{ns}$, but also in $\psi_\mathrm{r}$ implies uGMRT has a stable polarization response within an observation. 
Incorrect $\mathrm{D}_\mathrm{ns}$ directly impacts RM measurements and unstable $\psi_\mathrm{r}$ affects PA measurements. 
Also see Figs.~B1 and B2 of \citetalias{24BethapudiRM}.

\begin{figure}
    \centering
    \includegraphics[width=0.45\textwidth,keepaspectratio]{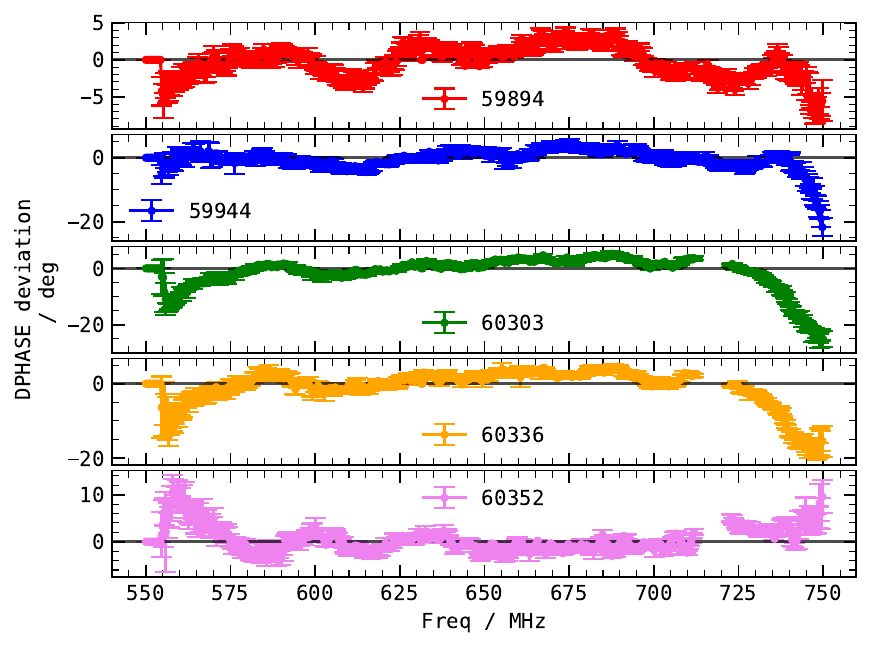}
    \caption{Deviations from linear \cmd{dphase} modeling shown against frequency using noise diode scans taken on different MJDs.
    Each panel shows deviation where each \cmd{dphase} fitted model is subtracted from the \cmd{dphase} data.
    }
    \label{fig:dpadev}
\end{figure}

\par The linear \cmd{dphase} modeling is simply a model. 
In every panel of Fig.~\ref{fig:dpadev}, we show the resultant when linear \cmd{dphase} fitted model is subtracted from data\cmd{dphase} for all the noise diode scans we observed in the specific MJD.
Since we have multiple noise diode scans, we show the mean and standard deviation computed for every frequency channel with errorbars.
We notice departures from the linear model to be of the order of $\pm 15$ deg.
We also note the following: (i) the deviations are consistent from scan to scan, and (ii) every observation has different form of deviation.
As mentioned in the main text, the impact of these unmodeled deviations on the resultant PA is insignificant. When averaged over the entire band, these deviations only change the averaged PA by order of 1 deg.

\begin{figure}
    \centering
    \includegraphics[width=0.45\textwidth,keepaspectratio]{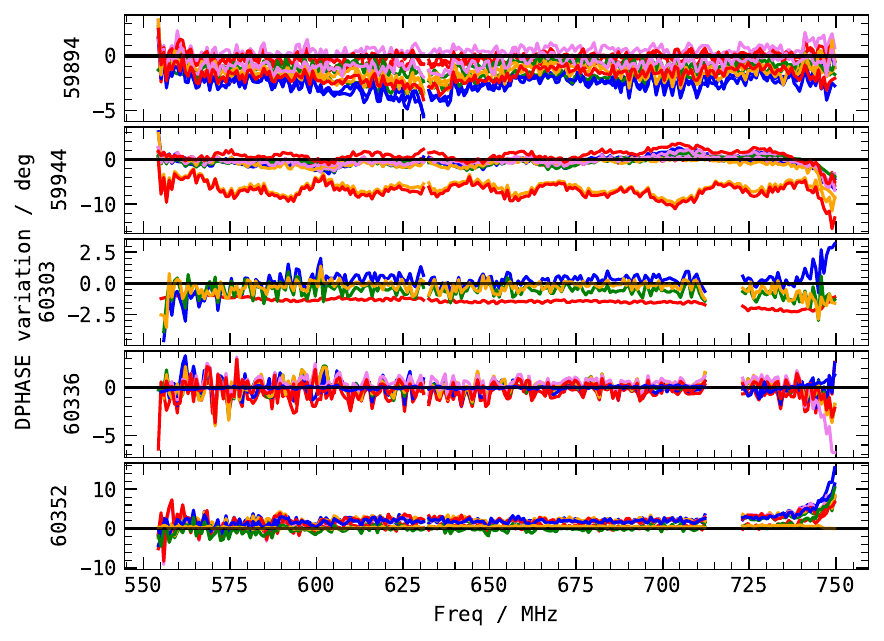}
    \caption{\cmd{dphase} variations from scan to scan shown against frequency for different MJDs.
    Each panel shows the difference of \cmd{dphase} between the first noise diode scan and the $n^\mathrm{th}$ scan for every observation.    
    }
    \label{fig:dpavar}
\end{figure}

\par Even though linear model and deviations from the linear model are consistent from one noise diode scan to another, \cmd{dphase} variations from one noise diode scan to another noise diode scan are of the order of $\pm 10$ deg.
To illustrate this, we subtract the \cmd{dphase} data of the first noise diode scan from that of all the remaining noise diodes scans.
We show the resultant against frequency for every observation in Fig.~\ref{fig:dpavar}.
We note the following: (i) variations are persistent in all the scans, as seen in MJD 60352 at 750 MHz, (ii) variations are drastic as seen in MJD 59944.

\section{Plots}
\label{a:sec:plots}

\begin{figure}
    \centering
    \includegraphics[width=0.45\textwidth,keepaspectratio]{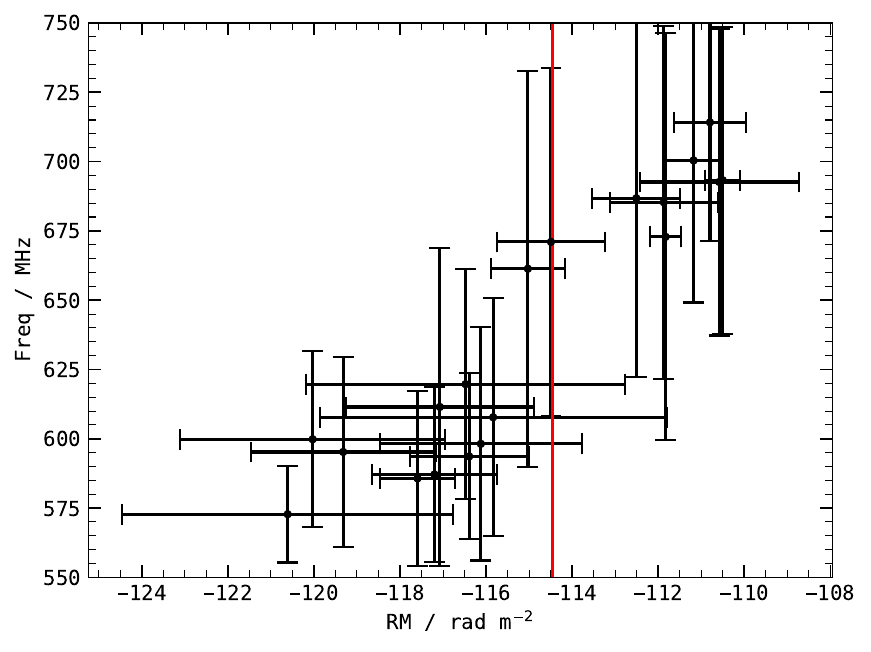}
    \caption{Bandwidth versus measured RM for a selection of bursts detected on MJD~59243. The bandwidth is shown as error in y-axis, while x-axis error refers to RM error. The red line is the inverse variance weighted RM average. Bursts have been selected such that they occupy around half of the entire observing band. The observed trend is non physical and caused by \cmd{dphase} deviations from the linear model used to calibrate. See Sect.~\ref{ssec:mrm}.}
    \label{fig:rmfreq}
\end{figure}

\begin{figure*}
    \centering
    \includegraphics[width=0.9\textwidth,keepaspectratio]{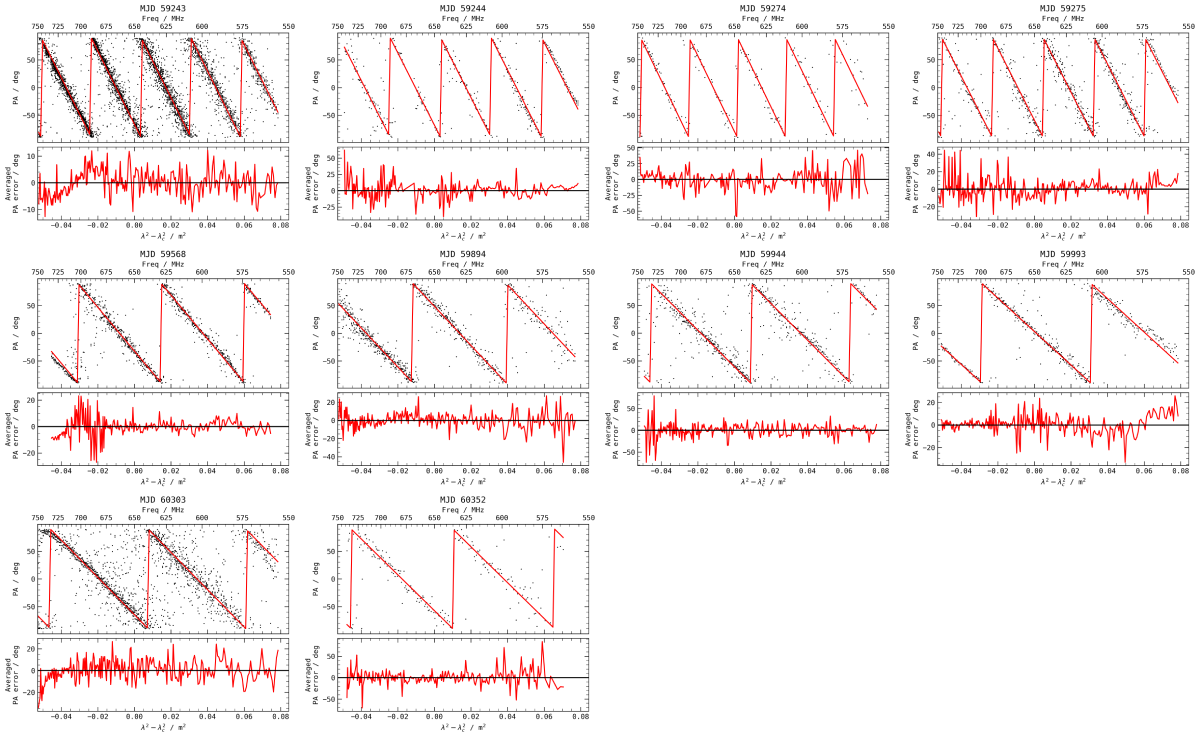}
    \caption{
    Outcomes of fitting one RM and one PA to all the bursts from an observation. Each plot corresponds to one observation. The top panels show PA against $\lambda^2-\lambda_c^2$ where $\lambda_c$ is the wavelength of the center frequency of the band. The individual data points are shown with black dots and the red line corresponds to the fitted model.
    The top x-axis shows frequency in MHz.
    The bottom panels show averaged error between data and fitted model. The horizontal black line is drawn at 0 deg.
    Individual high resolution plots are available separately. See Data availability.
    }
    \label{fig:fitonerm}
\end{figure*}

\begin{figure}
    \centering
    \includegraphics[width=0.45\textwidth,keepaspectratio]{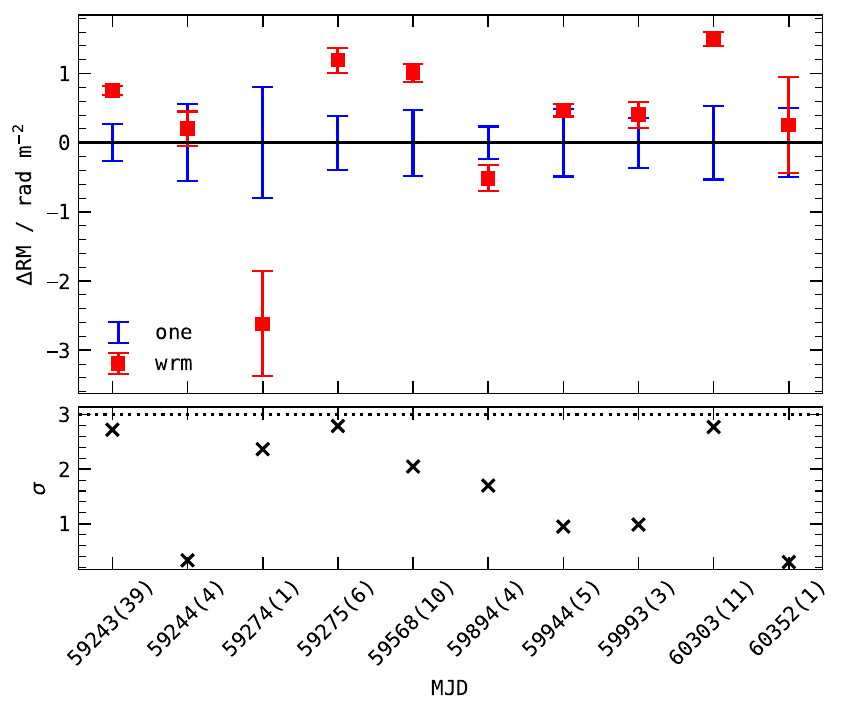}
    \caption{Difference between one RM fitted to all bursts in an observation (shown with \texttt{one} as legend) and the inverse variance weighted average of each individual $QU-$fitted RM measurements from each burst of the observation (shown with \texttt{wrm} as legend). The top panel only shows the difference with error-bars. The bottom panels shows the significance of difference. 
    The horizontal black dotted line is drawn at $3\sigma$.
    The number in parenthesis refers to the number of bursts detected in the respective MJD.}
    \label{fig:rmonewrm}
\end{figure}

\begin{figure}
    \centering
    \includegraphics[width=0.45\textwidth,keepaspectratio]{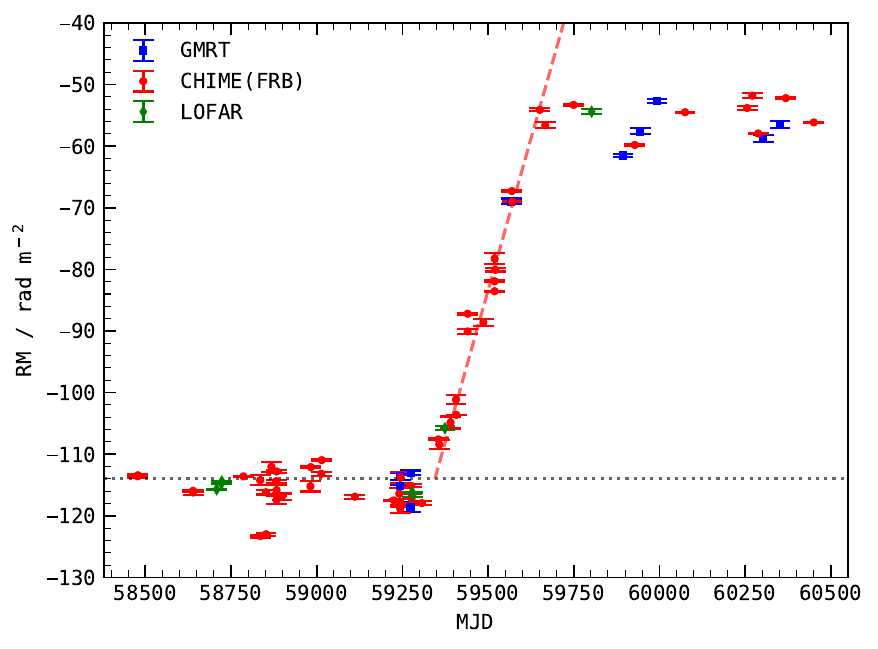}
    \caption{RM versus MJD. CHIME/FRB measurements \citep{23McKinvenR3,24NgRM} are shown in red circles and LOFAR measurements \citep{24GopinathR3} are shown in green diamonds. GMRT points, shown in blue squares, are the fitted RM values presented in Table~\ref{tab:ppdata}.}
    \label{fig:onermmjd}
\end{figure}

\begin{figure}
    \centering
    \includegraphics[width=0.45\textwidth, keepaspectratio]{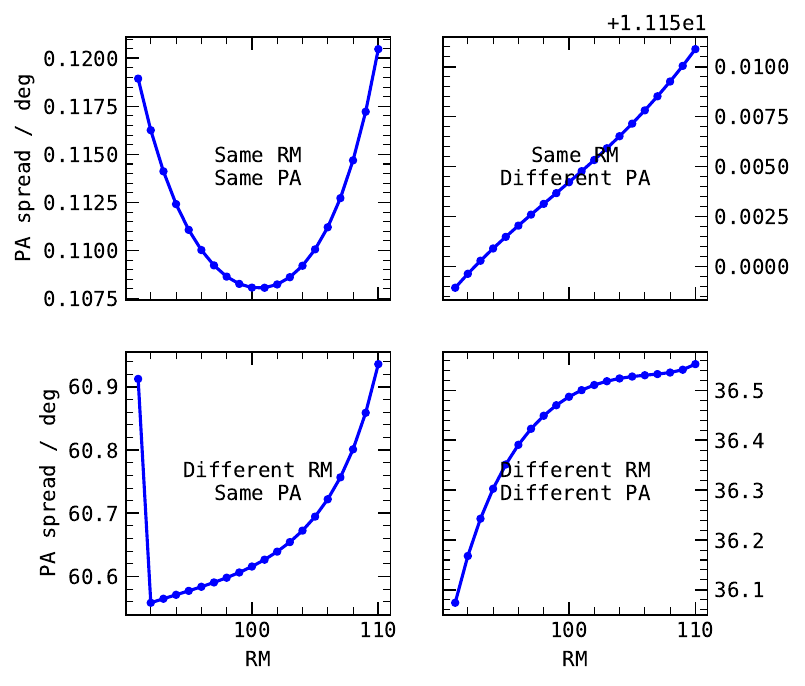}
    \caption{Showing by means of simulation that spread in PA is only minimized when bursts have non varying RM and PA. We simulate 15 bursts with same RM and same PA \emph{(top left)}, same RM but different PA \emph{(top right)}, different RM but same PA \emph{(bottom left)}, and different RM and different PA \emph{(bottom right)}. In each of the case, we correct for one RM and measure the spread in PA after correction, which is plotted with blue dots in each of the subplots. The spread in PA is greatly minimized and appears like a parabola only in the case of same RM and same PA.}
    \label{fig:paspreadgame}
\end{figure}

\begin{figure}
    \centering
    \includegraphics[width=0.45\textwidth,keepaspectratio]{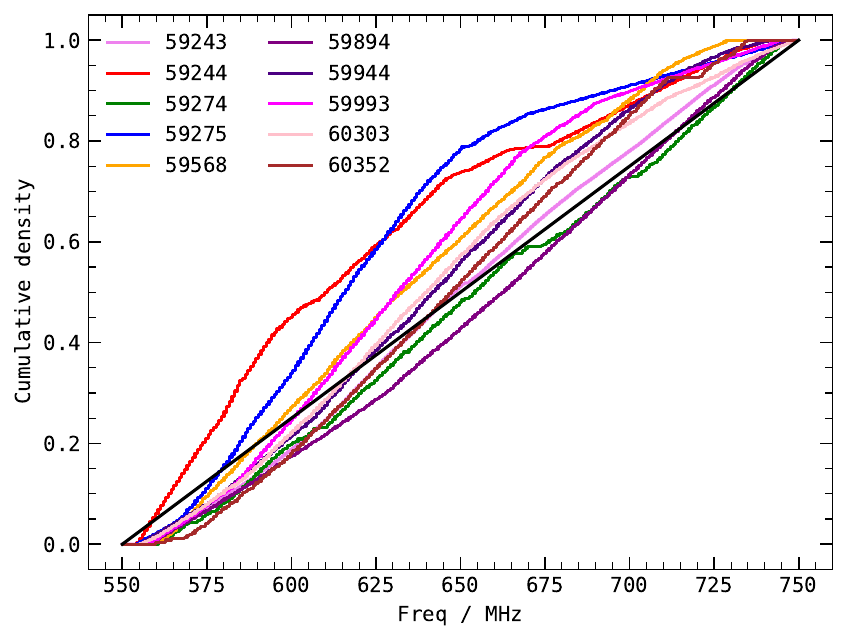}
    \caption{Cumulative distributions of the bursts of an observation over frequency band. The black bold line is the cumulative uniform distribution. Bursts from every observation occupy the whole observing band.}
    \label{fig:freqdist}
\end{figure}

\section{Slow precession simulation}
\label{sec:spsim}

\par In the inertial coordinate system defined in the main text (Sect.~\ref{ssec:slowp}, only the observer axis and the angular momentum axis are constants of motion.
The emission axis, however, varies with time in the inertial frame.
The angles between the projections of emission axis and angular momentum axis onto the plane perpendicular to the observer axis is defined as the PA.
Note that this requires the emission axis to be in the same inertial frame as the observer axis.
The projection of the angular momentum axis onto the same plane serves as a reference PA ($\mathrm{PA}_0$).
We bring the emission axis defined in the body-fixed frame using ($\alpha$ and $\phi_0$) into inertial frame using Euler angles ($\phi,\theta,\psi$) where the angles may or may not be functions of time.
The Euler angles are defined in ZXZ format.
The formalism here is similar to Sect.~37 of \citet{Landau1976Mechanics} and that used in \citet{24HeylIXPE}.
The $\theta = \delta$, that is, it is the spin misalignment angle. $\phi$ and $\psi$ track the rotational and precessional phase longitudes respectively.
Therefore, are defined as the following:
\begin{align*}
    \phi &= 2\pi\frac{t}{P_\mathrm{spin}} + \phi_0 \\
    \psi &= 2\pi\frac{t}{P_\mathrm{pre}},
\end{align*} where $t$ is the time and $P_\mathrm{spin}$ and $P_\mathrm{pre}$ are the spin and precessional periods respectively.
Note how $\phi_0$ can be re-defined as the initial value of the $\phi$ as both are degenerate.
There may also be a precessional phase offset which we have omitted here for simplicity.

\par
\begin{figure}
    \centering
    \includegraphics[width=0.45\textwidth,keepaspectratio]{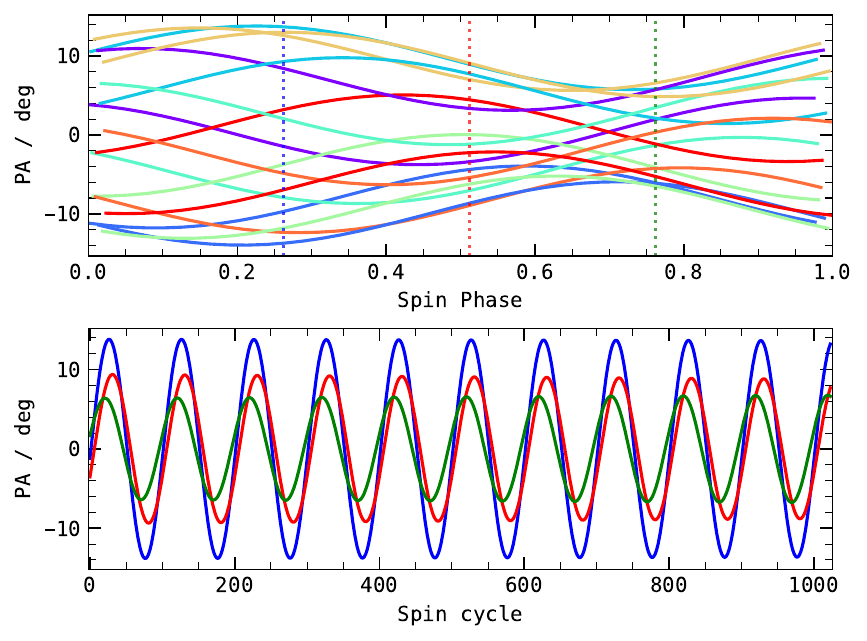}
    \caption{PAs measured using a mock model of slow precession. The parameters of the models are as follows: $\delta=10, \Psi_0=10, \alpha=4, \phi_0=200$ and $\zeta=90$ where all the angles are in degrees. \emph{Top} shows PA versus Spin phase for different Spin cycles using different colors. The three vertical dotted lines mark the Spin phases chosen to show PA versus Spin cycle in the \emph{bottom} panel. See Sect.~\ref{ssec:slowp}.}
    \label{fig:simslowp}
\end{figure}

 For the toy model, we set $\delta = 10$ deg, $\Psi_0 = 0$ deg, $\alpha = 4$ deg, $\phi_0 = 200$ deg and $\zeta = 90$ deg.
Moreover, we set spin period and precessional period to be $10$ and $1000$ days respectively.
Then, we measure the PAs for different spin phases and cycles.
We illustrate the PA versus spin phase for different spin cycles in Fig.~\ref{fig:simslowp} (\emph{top}).
The different colors imply different spin cycles.
We identify three unique spin phases (which we marked with vertical dotted lines of different colors) and plot the PA versus spin cycle at each of the spin phases in \emph{bottom} panel of the same figure.
We highlight that a slow precession model shows intra-cycle variability (variability over spin phase) and inter-cycle variability in the form of sinusoids (variability over spin cycle).
The period of the sinusoids is the same irrespective of the spin phase but the amplitude varies.

%\bsp	% typesetting comment
\label{lastpage}

\end{document}